\documentclass[times,preprint]{aastex631}

\usepackage[T1]{fontenc}
\usepackage{epsf,color,amsmath}

\usepackage{cancel}

\newcommand{\sfrac}[2]{\mathchoice%
  {\kern0em\raise.5ex\hbox{\the\scriptfont0 #1}\kern-.15em/
    \kern-.15em\lower.25ex\hbox{\the\scriptfont0 #2}}
  {\kern0em\raise.5ex\hbox{\the\scriptfont0 #1}\kern-.15em/
    \kern-.15em\lower.25ex\hbox{\the\scriptfont0 #2}}
  {\kern0em\raise.5ex\hbox{\the\scriptscriptfont0 #1}\kern-.2em/
    \kern-.15em\lower.25ex\hbox{\the\scriptscriptfont0 #2}} {#1\!/#2}}

\usepackage{comment}

\newcommand{\castro}{{\sf Castro}}

\newcommand{\maestroex}{{\sf MAESTROeX}}
\newcommand{\microphysics}{{\sf Microphysics}}

\newcommand{\amrex}{{\sf AMReX}}
\newcommand{\pynucastro}{{\sf pynucastro}}

\newcommand{\isot}[2]{$^{#2}\mathrm{#1}$}
\newcommand{\isotm}[2]{{}^{#2}\mathrm{#1}}

\newcommand{\gcc}{\mathrm{g~cm^{-3} }}
\newcommand{\cms}{\mathrm{cm~s^{-1} }}

\newcommand{\Ub}{\mathbf{U}}

\newcommand{\omegadot}{\dot{\omega}}
\newcommand{\Sdot}{\dot{S}}
\newcommand{\ddx}[1]{{\frac{{\partial#1}}{\partial x}}}

\newcommand{\ddt}[1]{{\frac{{\partial#1}}{\partial t}}}
\newcommand{\odt}[1]{{\frac{{d#1}}{dt}}}

\newcommand{\Ic}{{\boldsymbol{\mathcal{I}}}}

\usepackage{bm}

\newcommand{\Uc}{{\,\bm{\mathcal{U}}}}
\newcommand{\Fb}{\mathbf{F}}
\newcommand{\Sc}{\mathbf{S}}

\newcommand{\xv}{{(x)}}

\newcommand{\Ab}{{\bf A}}

\newcommand{\qb}{{\bf q}}

\newcommand{\Rb}{{\bf R}}

\newcommand{\Adv}[1]{{\left [\boldsymbol{\mathcal{A}} \left(#1\right)\right]}}

\newcommand{\Advss}[1]{{\left [{\mathcal{{A}}} \left(#1\right)\right]}}

\newcommand{\Advs}[1]{\boldsymbol{\mathcal{A}} \left(#1\right)}

\newcommand{\rhonse}{{\rho_\mathrm{nse}}}
\newcommand{\tnse}{{T_\mathrm{nse}}}
\newcommand{\Anse}{{A_\mathrm{nse}}}
\newcommand{\Bnse}{{B_\mathrm{nse}}}

\newcommand{\epsthermal}{\epsilon_{\nu,\mathrm{thermal}}}
\newcommand{\epsreact}{\epsilon_{\nu,\mathrm{react}}}

\newcommand{\bea}{\langle B/A \rangle}
\newcommand{\beav}{\left \langle \frac{B}{A} \right \rangle}
\newcommand{\dbeadt}{\left [ \frac{d\bea}{dt}\right]}
\newcommand{\yedotv}{\frac{dY_e}{dt}}
\newcommand{\yedotvp}{\left [ \frac{dY_e}{dt} \right ]}

\newcommand{\nsetable}{{\tt nse\_table}}
\newcommand{\dbeaweak}{(d\bea / dt)_\mathrm{weak}}
\begin{document}
%======================================================================
% Title
%======================================================================
%\title{A Simplified Spectral Deferred Correction Method for Coupling Hydrodynamics with Reaction Networks and Nuclear Statistical Equilibrium}
\title{Strong Coupling of Hydrodynamics and Reactions in Nuclear Statistical Equilibrium for Modeling Convection in Massive Stars}

%\shorttitle{A Simplified SDC Method}
\shorttitle{Hydro/Reaction Coupling with NSE}

\shortauthors{}

\author[0000-0001-8401-030X]{Michael Zingale}
\affiliation{Department of Physics and Astronomy, Stony Brook University, Stony Brook, NY 11794-3800, USA}

\author[0000-0002-2839-107X]{Zhi Chen}
\affiliation{Department of Physics and Astronomy,
Stony Brook University,
Stony Brook, NY 11794-3800, USA}

\author[0000-0003-3603-6868]{Eric T. Johnson}
\affiliation{Department of Physics and Astronomy,
Stony Brook University,
Stony Brook, NY 11794-3800, USA}

\author[0000-0003-0439-4556]{Max P.~Katz}
\affiliation{Department of Physics and Astronomy, Stony Brook University, Stony Brook, NY 11794-3800, USA}

\author[0000-0001-5961-1680]{Alexander Smith Clark}
\affiliation{Department of Physics and Astronomy,
Stony Brook University,
Stony Brook, NY 11794-3800, USA}

\correspondingauthor{Michael Zingale}
\email{michael.zingale@stonybrook.edu}

%======================================================================
% Abstract and Keywords
%======================================================================
\begin{abstract}
We build on the simplified spectral deferred corrections (SDC) coupling of
hydrodynamics and reactions to handle the case of nuclear statistical
equilibrium (NSE) and electron/positron captures/decays in the cores of massive stars.  Our approach blends a traditional reaction network on the grid with a tabulated
NSE state from a very large, $\mathcal{O}(100)$ nuclei, network.  We demonstrate how to achieve
second-order accuracy in the simplified-SDC framework when coupling NSE to hydrodynamics, with the ability to evolve the star on the hydrodynamics timestep.  We discuss the application of this method to convection in massive stars leading
up to core-collapse.  We also show how to initialize the initial convective state from a 1D model in a self-consistent fashion.  All of these developments
are done in the publicly available \castro\ simulation code and the
entire simulation methodology is 
fully GPU accelerated.
\end{abstract}


%======================================================================
% Introduction
%======================================================================
\section{Introduction}\label{Sec:Introduction}

When modeling reactive flows in stars, the energy release
from nuclear reactions drives a variety of complex behaviors, including convection, flames, and detonations.  In massive stars, during the final approach to a core-collapse
supernova, capturing the energy release can be especially challenging as the Si burning is very energetic
and, in the core, a large number of electron-capture reactions need to be accounted for.  This demands a large reaction network, which is computationally expensive in multi-dimensions.  The direct feedback between the reactions and the hydrodynamics means that care needs to be taken to ensure that the reactions and hydrodynamics are properly coupled together.  The difficulty in doing this stems from the fact that explicit-in-time integration is usually preferred for hydrodynamics while implicit methods are usually needed for reactions due to their stiff nature.  Traditionally, an operator splitting approach has been used to update the fluid state in time.

One-dimensional models of massive star evolution leading to core-collapse can use hundreds of nuclei (e.g. \citealt{rauscher:2002})  including important electron/positron captures and decays \citep{heger:2001}, allowing for detailed evolution of the electron fraction in the iron core.
In contrast, most multidimensional models of massive star convection use smaller networks, like the 21-nuclei {\tt aprox21} network, as their workhorse. These smaller networks may not be able to capture the energetics and nucleosynthesis \citep{farmer:2016,kato:2020,navo:2023}.

Multidimensional models of massive star evolution have explored a variety of progenitor masses, focusing on different aspects of the evolution, with a range of different algorithms, including differences in the grid geometry, evolution of the core, and reaction network (and its integration).  Here we summarize these differences, focusing on the geometry and treatment of the core.

The work of \citet{arnettmeakin:2011} considered  
2D simulations of the C, Ne, O, and Si shells, using a reaction network with 38 species. They showed that large asymmetries in the burning shells can form, in contrast to the classic ``onion skin'' picture from one-dimensional models.   This work used spherical coordinates and cut out the inner iron core.

\citet{couch:2015,chatzopoulos:2016,fields:2020,fields:2021} looked at the late stages of convection
leading up to core collapse in 3D with a variety of progenitors,
using the {\tt aprox21} network. These all used a Cartesian geometry, with differing treatments of the iron core.  In \citet{fields:2020,fields:2021}, the evolution of the core was done by interpolating the evolution from the one-dimensional MESA model, acting as an effective inner boundary condition.  Due to the operator splitting of hydrodynamics and reactions, they also used a reaction-based timestep, only allowing the internal energy to change by 1\% per timestep.

Other recent work used 3D spherical geometry with the core not modeled  \citep{muller:2017,varmamueller:2023}, or 3D with the core evolved in 1D with spherical symmetry \citep{yoshida:2019,yoshida-apj:2021,yoshida:2021}.
The {\tt aprox21} network in \citet{yoshida-apj:2021} transitioned to nuclear statistical equilibrium (NSE) at high temperatures, similar to the approach that we will discuss here.  An alternate 3D yin-yang geometry was used by \citet{yadav:2020}, again with the core  cutout.

All of these works have shown that 
asymmetries can have important impacts on the core-collapse explosion mechanism (see also \citealt{muller:2020}).  Capturing the evolution
of the core is difficult however, both due to the coordinate singularity in some discretizations as well as the differing timescales of the reactions in the core versus the convective shells.  Additionally, many weak interactions are important in the evolution of the electron fraction in the core that cannot be captured with a simple network (see, e.g., \citealt{farmer:2016}).

Our goal here is to show how to model a massive star on the hydrodynamics timescale, instead of restricting the timestep based on the energy release, by using strongly-coupled time-integration strategies.  We also aim to self-consistently evolve the core by capturing the NSE state, and show how to do this with second-order accuracy.  Finally, we show how to establish an initial convective velocity field from a 1D model.
Since our focus here is on the time-integration algorithm, we work in a 2D axisymmetric geometry.

In \cite{castro_simple_sdc} (henceforth Z22), we introduced a coupling
method for hydrodynamics and reactions based on the ideas of spectral
deferred corrections (SDC) \citep{dutt:2000,minion:2003,castro_sdc},
but designed to work with the standard characteristic-tracing approach
to hydrodynamics used in the corner transport upwind (CTU)
\citep{ppmunsplit} method and its extension to the piecewise parabolic
method (PPM) \citep{ppm,millercolella:2002}.  We termed this
integration method ``simplified-SDC''.  With operator splitting, when
the reactions are evolved, the expansion of the fluid via mechanical
work is not seen.  This means that if a lot of heat is released into
  a zone in a timestep, we will not see the expansion and cooling that
  should accompany this.  In contrast to operator splitting, in SDC
  the advection and reactions explicitly see the effects of each
  other, and in Z22, we showed that this allows us to model explosive
  reactive flows more accurately and efficiently.

\section{Numerical Methodology}

In this section, we will describe how to augment the equations of reacting
hydrodynamic flow to accommodate the composition variables needed from NSE, and perform the coupled time-integration of the system.  We use the \castro\ compressible astrophysics simulation code \citep{castro,castro_joss} together with the \amrex\ adaptive mesh refinement library \citep{amrex_joss}.
We will follow the notation from Z22 and the composition
approach of \citet{ma:2013}.  For simplicity,
we write things out for a one-dimensional system, but the extension to
multi-dimensions is straightforward.  

The conservative-form of the one-dimensional
reactive Euler equations can be written as:
\begin{equation}
  \ddt{\Uc} + \ddx{\Fb(\Uc)} = \Sc(\Uc)
\end{equation}
with the conserved state
\begin{equation}
  \Uc = \left ( \rho, \rho X_k, \rho \alpha_l, \rho u, \rho E, \rho e\right )^\intercal
\end{equation}
where $\rho$ is the mass density, $u$ is velocity, $E$ is specific
total energy, $e$ is the specific internal
energy, $X_k$ are the mass fractions,
subject to $\sum_k X_k = 1$, and $\alpha_l$ are auxiliary
composition variables.  Following \citet{ma:2013}, we carry three
auxiliary composition variables: electron fraction, $Y_e$, 
\begin{equation}
  \label{eq:aux:ye}
  Y_e = \sum_k \frac{Z_k X_k}{A_k}
\end{equation}
where $A_k$ and $Z_k$ are the atomic weight and atomic number of nucleus $k$,
mean molecular weight, $\bar{A}$,
\begin{equation}
\label{eq:aux:abar}
\bar{A} = \left [ \sum_k \frac{X_k}{A_k} \right ]^{-1} \enskip ,\\
\end{equation}
and average binding energy per nucleon, $\langle B/A\rangle$:  
\begin{equation}
\label{eq:aux:bea}
\left \langle \frac{B}{A} \right \rangle = \sum_k \frac{B_k X_k}{A_k} \enskip ,
\end{equation}
where $B_k$ is the binding energy of nucleus $k$.  
These same composition variables are used in the NSE implementation of \citet{townsley:2007}.
Finally, the specific total energy and specific
internal energy are related via:
\begin{equation}
E = e + u^2/2
\end{equation}

The corresponding hydrodynamical fluxes are:
\begin{equation}
  \Fb(\Uc) = \left ( \begin{array}{c}
         \rho u \\
         \rho X_k u \\
         \rho \alpha_l u \\
         \rho u^2  + p \\
         (\rho E + p) u \\
         \rho e u \end{array}\right )
\end{equation}
where $p$ is the pressure.
We 
separately evolve $\rho e$, as part of a dual energy formalism \citep{bryan:1995,wdmergerI},
and $(\rho e)$ evolution equation includes an additional ``pdV'' term that accounts for the variation of internal energy due to compression of the fluid element:
\begin{equation}
\ddt{(\rho e)} + \ddx{F(\rho e)} + p \ddx{u} = S_{\rho e}
\end{equation}

Again following \cite{castro_simple_sdc}, we split the source term into hydrodynamic and reactive parts:
\begin{equation}
  {\bf G}(\Uc) = \left ( \begin{array}{c}
    0 \\
    0 \\
    0 \\
    \rho g \\
    \rho u g \\
    0 \end{array} \right )
  \qquad
  \Rb(\Uc) = \left ( \begin{array}{c}
     0 \\
     \rho \omegadot_k \\
     \rho \omegadot_l \\
     0 \\
     \rho \epsilon \\
     \rho \epsilon 
  \end{array} \right )
\end{equation}
with 
\begin{equation}
  \Sc(\Uc) = {\bf G}(\Uc) + \Rb(\Uc).
\end{equation}
where $g$ is the gravitational acceleration.
Here, 
$\omegadot_k$ is the creation rate for species $k$,
$\omegadot_l$ is the creation rate for auxiliary composition
variable $l$, and $\epsilon$ is the energy generation rate per unit
mass (we labeled this $\Sdot$ in Z22).  We note that although we set $G_{\rho e}$ and $G_{\rho E}$ to zero for the current application, it could also include conductive radiation-transport terms.  These reactive sources all depend on density, temperature, and composition.  Finally, we define an advective term including the hydrodynamic sources:
\begin{equation}
\Advs{\Uc} = -\ddx{\Fb(\Uc)} + {\bf G}(\Uc)
\end{equation}
and again, the $(\rho e)$ component is treated specially:
\begin{equation}
\mathcal{A}(\rho e) = -\ddx{F(\rho e)} -p \ddx{u} + G_{\rho e}
\end{equation}

The equation of state we use in this work takes a slightly different algebraic
form than that in Z22 because of how
we define the composition.  For a system
where the composition is completely specified by the mass fractions,
$X_k$, the equation of state would take the form:
\begin{equation}
p = p(\rho, X_k, e); \; T = T(\rho, X_k, e)
\end{equation}
In this work, we will instead use the auxiliary composition
variables, $\alpha_l$, and the EOS has the form:
\begin{equation}
p = p(\rho, \alpha_l, e); \; T = T(\rho, \alpha_l, e)
\end{equation}
We use the Helmholtz stellar equation of state of \citet{timmes_swesty:2000,flash} for all simulations here.

Our primitive variable system is also slightly different than that
of Z22 because of the $\alpha_l$ term.  The 
primitive variables are:
\begin{equation}
\qb = \left ( \rho, X_k, \alpha_l, u, p, (\rho e) \right )^\intercal
\end{equation}
and writing the system as:
\begin{equation}
\qb_t + \Ab^\xv(\qb) \qb_x  = \Sc(\qb)
\end{equation}
we have the matrix $\Ab^\xv$:
\begin{equation}
\Ab^\xv(\qb) = \left ( \begin{array}{cccccc}
    u & 0 & 0 & \rho & 0 & 0 \\
    0 & u & 0 & 0    & 0 & 0 \\
    0 & 0 & u & 0    & 0 & 0 \\
    0 & 0 & 0 & u    & 1/\rho & 0 \\
    0 & 0 & 0 & \Gamma_1 p & u & 0 \\
    0 & 0 & 0 & \rho h & 0 & u
  \end{array} \right )
\end{equation}
where $h$ is the specific enthalpy and $\Gamma_1$ is an adiabatic index,
$\Gamma_1 = d\log p/d\log\rho|_s$ at constant entropy.  This form
of the equations is used in the construction of the interface states
and the characteristic decomposition of the jumps that reach the interface
over the timestep. Similar to conservative variables we
decompose the source terms, $\Sc(\qb)$ as:
\begin{equation}
  \Sc(\qb) = {\bf G}(\qb) + \Rb(\qb)
\end{equation}
with
\begin{equation}
\label{eq:prim_sources}
{\bf G}(\qb) = \left ( \begin{array}{c}
     0 \\
     0 \\
     0 \\
     g \\
     0 \\
     0 \\
   \end{array} \right )
\qquad
\Rb(\qb) = \left ( \begin{array}{c}
     0 \\
     \omegadot(X_k) \\
     \omegadot(\alpha_l) \\
     0 \\
     \Gamma_1 p \sigma \epsilon \\
     \rho \epsilon
   \end{array} \right )
\end{equation}
where $\sigma$ is
\begin{equation}
\sigma \equiv \frac{\partial p/\partial T |_\rho}{\rho c_p \partial p/\partial \rho |_T}
\end{equation}
with $c_p$ as the specific heat at constant pressure, $c_p = \partial
h/\partial T |_p$. (see \citealt{ABNZ:III}).

\subsection{Overview of the simplified-SDC approach}

The main idea of the simplified-SDC approach is that the hydrodynamics update explicitly
sees a reactive source term and that the reaction network integration includes an advective term---this strongly
couples the two processes and avoids the splitting error seen in the commonly used Strang operator splitting 
approach \citep{strang:1968,strang_rnaas}.
The full algorithm is described in Z22.  
The overall update is iterative, so we use the superscript $(k)$ below
to denote the iteration.  The essential steps of the update are:
\begin{itemize}
  \item {\em Create the advective update term, $\Adv{\Uc}^{n+1/2,(k)}$}, using the CTU PPM method \citep{ppmunsplit,millercolella:2002}.
    The interface states will see a source term in the prediction step that takes the form:
      \begin{equation}
        \Sc(\qb) = {\bf G}(\qb) + \Ic_\qb^{n+1/2,(k-1)}
      \end{equation}
    where $\Ic_\qb^{n+1/2,(k-1)}$ is an approximation to the integral
    of the reaction sources over the timestep.  In the first
    iteration, this is the source term from the previous step, in
    subsequent iterations, it is the source from the previous
    iteration.

  \item {\em Update the system using an ODE integrator.}
    We update the entire advection-reaction system as:
    \begin{equation}
      \odt{\Uc} = \Adv{\Uc}^{n+1/2,(k)} + \Rb(\Uc)
    \end{equation}
    where we take the advective term to be piecewise constant in time---this
    makes the system an ODE system, and can be solved using a standard ODE integrator.  Since $\Adv{\Uc}^{n+1/2,(k)}$ is time-centered, this is
    second-order accurate as long as we use a high-order integrator for the reaction ODE system.

    In practice, we only need to evolve the subsystem
    \begin{equation}
      \Uc^\prime = \left ( \rho X_k, \rho \alpha_l, \rho e \right )^\intercal
    \end{equation}
    and integrate
    \begin{equation}
      \odt{\Uc^\prime} = \Adv{\Uc^\prime}^{n+1/2,(k)} + \Rb(\Uc^\prime) \enskip , \label{eq:reaction_ode}
    \end{equation}
    since they are the only quantities that depend on reactions.  The remaining
    quantities can simply be updated with the advective term, e.g.,
    \begin{equation}
        \rho(t) = \rho^n + (t - t^n) \Advss{\rho}^{n+1/2,(k)}
    \end{equation}
    Furthermore, when integrating the reaction network directly, outside of NSE, we
    do not evolve $\alpha_l$, but instead compute them as needed from the mass fractions.
    This system is evolved using the VODE ODE integrator \citep{vode}, with the modifications
    described in \cite{castro_simple_sdc}, although several
    other integrators are available in our \microphysics\ suite (see, e.g., \citealt{rkc_rnaas}).
    
    This integration begins with $\Uc^{n}$ and results in $\Uc^{n+1,(k)}$.

  \item {\em Compute the reactive source terms.}
    Finally, we compute the $\Ic$'s that capture the effect of just the
    reaction sources on the primitive state variables for the next iteration.
    This takes the form 
    \begin{equation}
      \label{eq:Iq}
      \Ic^{(k)}_{\qb} = \frac{\qb^{n+1,(k)} - \qb^n}{\Delta t} - \Adv{\qb}^{n+1/2,(k)}
    \end{equation}
    and we detailed how to compute this in Z22.
    We note that in the current work, this includes a source term for
    the $\alpha_l$, $\Ic_q^{(k)}(\alpha_l)$.
\end{itemize}

Two iterations are required for second order convergence in time.

\section{Reaction Networks and Nuclear Statistical Equilibrium}

Since we are using the auxiliary composition, $\alpha_l$, to specify the 
thermodynamics, we included an advection equation for them above.  
We can derive evolution equations for each of these composition quantities as:
\begin{eqnarray}
\frac{DY_e}{Dt} &=& \sum_k \frac{Z_k}{A_k} \frac{DX_k}{Dt} = \sum_k \frac{Z_k}{A_k} \omegadot_k \label{eq:ye_evolve}\\
\frac{D\bar{A}}{Dt} &=& -\bar{A}^2 \sum_k \frac{1}{A_k} \frac{DX_k}{Dt} = -\bar{A}^2 \sum_k \frac{1}{A_k} \omegadot_k \\
\frac{D}{Dt} \left \langle \frac{B}{A} \right \rangle &=& \sum_k \frac{B_k}{A_k} \frac{DX_k}{Dt} = \sum_k \frac{B_k}{A_k} \omegadot_k \label{eq:beaevolve}
\end{eqnarray}
For Strang split coupling of hydro and reactions, $\omegadot_k = DX_k/Dt = 0$,
when doing the hydro update, and 
therefore each of these auxiliary quantities, $\alpha_l$, obeys an advection
equation in the hydro part of the advancement:
\begin{equation}
\frac{\partial (\rho \alpha_l)}{\partial t} + \nabla \cdot (\rho \Ub \alpha_l) = 0
\end{equation}
However, in the SDC approach, we will include the $\Ic_q(\alpha_l)$ term in the prediction
of the interface states to the Riemann solver that
captures the source term in Eqs.~\ref{eq:ye_evolve}--\ref{eq:beaevolve}.

\subsection{Reaction Network and NSE}

For this study, we use a reaction network with 19 nuclei:
\isot{H}{1}, \isot{He}{3}, \isot{He}{4}, \isot{C}{12}, \isot{N}{14},
\isot{O}{16}, \isot{Ne}{20}, \isot{Mg}{24}, \isot{Si}{28},
\isot{S}{32}, \isot{Ar}{36}, \isot{Ca}{40}, \isot{Ti}{44},
\isot{Cr}{48}, \isot{Fe}{52}, \isot{Fe}{54}, \isot{Ni}{56}, protons
(from photodisintegration), and neutrons.  This is based on the
{\tt aprox19} originally described in
\cite{Kepler}.  This describes H and He burning well, and builds an alpha-chain consisting of $(\alpha,\gamma)$ and $(\alpha,p)(p,\gamma)$ reactions up to \isot{Ni}{56} (as the product of Si-burning).  The slightly larger {\tt aprox21} network is often used for massive stars, since it includes \isot{Fe}{45} and \isot{Cr}{48}, allowing the network to reach lower $Y_e$.  But it does not include all of the electron/positron captures and decays that will be taking place in the core of the massive star.  To account for these, we blend our
network with a table of NSE that provides the $Y_e$ evolution in the core.  NSE tables have been used for modeling Type Ia supernova, for example in \citet{townsley:2007,seitenzhal:2009,ma:2013}, with the latter using \castro\ and blending the NSE state with a 7-isotope network using simple Strang-split coupling.

For this work, we generate the NSE table following the approach of
\citet{seitenzhal:2009}, as implemented in
\pynucastro\ \citep{pynucastro2}.  Our goal with this paper is to
describe the framework for coupling NSE and reactions to hydrodynamics
in an accurate fashion.  In a follow-on work, we will explore how the
details of the nuclear physics used with the on-grid network and NSE
table affect the simulations.  The entire process for generating the
NSE table is open source, allowing for the table to easily be
updated with new rates.  By controlling the generation of the table
ourselves, we can experiment with more recent weak rate tabulations
and create custom mappings to the on-grid nuclei for a more extensive
network.  An overview of the table generation is given in Appendix
\ref{app:nse}.  Similar to other NSE tabulations, our NSE table takes
as inputs $\rho$, $T$, and $Y_e$ and tabulates $\bar{A}$, $dY_e/dt$,
and $\bea$, $\dbeaweak$, and $\epsreact$, where the ``weak'' subscript
indicates that these terms are due to the weak rate evolution of the
NSE state.  By storing $dY_e/dt$, $\dbeaweak$, and the weak rate
neutrino energy loss, $\epsreact$, we are able to capture the
evolution of the NSE state due to electron/positron captures and
decays over a timestep.  We note that $(d\bar{A}/dt)_\mathrm{weak}$ is
very small, so we do not consider it.

In our simulations, we carry all 19 isotopes in the main network in
each zone and advect them in the hydrodynamics portion of the
algorithm.  The 96 nuclei in the NSE table are mapped into the 19
isotopes we carry which are also stored on the table.  We'll refer to
this reduced composition as $\tilde{X}_k$.  Because the NSE table uses
many more nuclei, these approximate $\tilde{X}_k$ do not satisfy
Eqs.~\ref{eq:aux:ye} to \ref{eq:aux:bea} directly.  This is why we
need to explicitly carry the $\alpha_l$ on the grid and evolve them in
the hydrodynamics system.

Overall, the NSE function takes the form:
\begin{equation}
\nsetable(\rho,T,Y_e) \rightarrow \bar{A}, \tilde{X}_k, \beav, \yedotv, \dbeadt_\mathrm{weak}, \epsreact
\end{equation}

We found that to get second-order convergence, we need to use tricubic
interpolation with each direction equivalent to a cubic Lagrange
polynomial.  Ultimately, the uncertainty with which nuclear
masses are measured experimentally \citep{ame2016} may limit
the convergence.

Our equation of state needs the mean charge per nucleus, $\bar{Z}$, in addition
to the auxiliary quantities, which is computed as
\begin{equation}
\bar{Z} = \bar{A} \sum_k \frac{Z_k X_k}{A_k} = \bar{A} Y_e
\end{equation}
(see, e.g., \citealt{flash}).

\subsubsection{Initialization}

For any problem setup, we need to initial the $\alpha_l$.
We first initialize the mass fractions, $X_k$, and $Y_e$, by mapping from a one-dimensional initial model (or using a simple analytic expression for test problems).  If $Y_e$ is not available, then we compute $Y_e$ using (\ref{eq:aux:ye}).
For the two other composition
quantities we carry, $\bar{A}$, and $\bea$, we need values that are
consistent with the value of $Y_e$ and the nuclei stored in the NSE
table.  Therefore, if the thermodynamic conditions put us in NSE
(using the conditions defined below), then we obtain $\bar{A}$ and
$\bea$ from the NSE table, and use the corresponding $\tilde{X}_k$ instead of
the $X_k$ from the initial model. Otherwise, we compute these directly from
$X_k$ using (\ref{eq:aux:abar}) and (\ref{eq:aux:bea}).

\subsection{NSE condition}

For each zone, we need to be able to determine if we are in NSE.
Since we integrate energy during the reactions (see
Eq.~\ref{eq:reaction_ode}) we may begin a zone update out of NSE but
enter NSE in the middle of a timestep as the temperature rises. We
will address this possibility in the next section.  Since we don't
have the capability to evaluate individual rates from the table, we
use heuristics to determine if NSE applies.  We treat a zone as being
in NSE if the density and temperature exceed a threshold and the
composition is mainly $\alpha$ and Fe-group nuclei (\isot{Cr}{48},
\isot{Fe}{52}, \isot{Fe}{54}, and \isot{Ni}{56}).  The full condition
is:
\begin{eqnarray}
\rho &>& \rhonse \label{eq:nse_start}\\
T &>& \tnse \\
X(\isotm{He}{4}) + \sum_{k \in \mathrm{Fe\mbox{-}group}} X_k &>& \Anse \\
X(\isotm{Si}{28}) &<& \Bnse  \label{eq:nse_end}
\end{eqnarray}
Typical values are $\rhonse = 10^7~\gcc$, $T_\mathrm{nse} =
3\times 10^9~\mathrm{K}$, $\Anse=
0.88$, and $\Bnse = 0.02$.  The value of $\Bnse$ ensures that Si-burning has largely ended before invoking the table.  The value of $\Anse$ is based on \citet{ma:2013} but our $\Bnse$ is slightly
larger due to the table differences---we need to make sure we are above the maximum amount of $\isotm{Si}{28}$ that the NSE state will have (which is $\tilde{X}(\isotm{Si}{28}) = 0.012$ for our table).

\section{Coupling NSE with Reaction Networks and Hydrodynamics}

We now discuss how to couple hydrodynamics and reactions that potentially enter NSE 
using the simplified-SDC method.  For our reactions,
there are 3 parts to the energy generation:
\begin{equation}
\epsilon = \epsilon_\mathrm{nuc} - \epsthermal - \epsreact
\end{equation}
Here $\epsilon_\mathrm{nuc}$ is the energy generation from the nuclear reactions, $\epsthermal$ is the sum of the plasma, photo-, pair-, recombination, and Bremsstrahlung neutrino losses computed via \citet{itoh:1996}, and, when using the NSE table, $\epsreact$ are the neutrino losses from electron-captures and beta-decays (and their positron equivalents).  Each of these terms needs to be dealt
with separately.

When integrating the network directly, we evolve
an energy equation of the form:
\begin{equation}
\frac{d(\rho e)}{dt} =  \Advss{\rho e}^{n+1/2,(k)} + \rho \epsilon_\mathrm{nuc} - \rho \epsthermal
\end{equation}
calling the equation of state to get the temperature each time it is needed and evaluating the rates using the {\tt aprox19} network, with no contribution from $\epsreact$.
We then compute $\epsilon_\mathrm{nuc}$ from the change in mass:
\begin{equation}
    \epsilon_\mathrm{nuc} = - N_A c^2 \sum_k  \frac{dY_k}{dt} m_k
\end{equation}
(see, e.g., \citealt{hixmeyer}) where $Y_k = X_k / A_k$ are the molar fractions of the species and $m_k$ are the nuclei masses, defined as:
\begin{equation}
    m_k = (A_k - Z_k) m_n + Z_k (m_p + m_e) - B_k / c^2
\end{equation}
where $m_n$ is the neutron mass, $m_p$ is the proton mass, and $m_e$ is the electron mass.

When using the NSE table, we carry the average binding energy per nucleon, so we compute the energy release as:
\begin{equation}
\epsilon_\mathrm{nuc} = N_A \frac{d \bea}{dt} + N_A \Delta m_{np} c^2 \frac{dY_e}{dt}
\end{equation}
where $\Delta m_{np} = m_n - (m_p + m_e)$ and $N_A$ is Avogadro's number and the second
term captures the energy release from weak interactions. This expression is consistent with
that shown in \citet{townsley:2007,seitenzhal:2009}.

\subsection{SDC-NSE Coupling}
\label{sec:sdc-nse}

With SDC evolution, when we are in NSE, we need to do the advective
and reactive updates together.  A second-order accurate update takes the form:
\begin{equation}
\Uc^{n+1,(k)} = \Uc^n + \Delta t \Adv{\Uc}^{n+1/2,(k)} + \Delta t \left [\Rb (\Uc) \right ]^{n+1/2,(k)}
\end{equation}
For density and momentum, we can do this update already, since there
are no reactive sources.  That gives us $\rho^{n+1,(k)}$ and $(\rho
u)^{n+1,(k)}$.  For the other quantities, we do an update based on second-order Runge-Kutta.

Care needs to be taken with regards to
the binding energy.
We come into the NSE update with a current value for
$[\rho \langle B/A\rangle]^n$.  If we were already in
NSE at the end of the previous timestep, then this should
be the same average binding energy of the fluid element as
the NSE table would give.  But if we are coming
from a zone that was previously not in NSE (either at the
last timestep or because we bailed out of the current step's
integration early) then this $\langle B/A\rangle$ is not consistent
with NSE.  This means that there are 2 contributions to the
energy release during the NSE update: the energy resulting from
the change in $\langle B/A\rangle$ from instantaneously converting
the input state to NSE and the energy resulting from the evolution
of the NSE state over $\Delta t$ due to weak reactions and the advective changes in the fluid state.  As long as we use the incoming $\bea$ as the starting point and ensure that $\bea$ on leaving is consistent with the NSE table, we capture both effects.

\subsection{EOS Consistency}

An additional subtlety comes into play when we try to 
invert the equation of state to get temperature from
energy, since in NSE, the
composition depends on temperature.  For our equation of state, we have:
\begin{equation}
e = e(\rho, T, \bar{A}, \bar{Z})
\end{equation}
but the NSE requirement means $\bar{A}$ and $\bar{Z}$ are also functions of the state, so we write this as:
\begin{equation}
e = e^\mathrm{NSE}(\rho, T, \bar{A}(\rho, T, Y_e), \bar{Z}(\rho, T, Y_e)) \label{eq:ense}
\end{equation}
We'll define the procedure of finding a temperature consistent with the EOS and NSE as ${\tt nse\_eos}$:
\begin{equation}
{\tt nse\_eos}(\rho, e, Y_e) \rightarrow T, \bar{A}
\end{equation}
Given an energy, $e^\star$, the EOS procedure is:
\begin{itemize}
  \item Pick a guess $T_0$ for the temperature.
  \item Use the NSE table with $T_0$ to get $\bar{A}$ and $\bar{Z} = Y_e \bar{A}$.
  \item Define the function we will zero:
  \begin{equation}
  f(T) = e^\mathrm{NSE}(\rho, T, \bar{A}, \bar{Z}) - e^\star
  \end{equation}
  with $\rho$ and $Y_e$ held constant.
  \item Define the correction via Newton-Raphson iteration as:
  \begin{equation}
  \delta T = - \frac{f(T_0)}{df/dT |_{T_0}}
  \end{equation}
  with
  \begin{equation}
    \left . \frac{\partial f}{\partial T} \right |_{T_0} =
       \left . \frac{\partial e}{\partial T} \right |_{\rho, \bar{A}, \bar{Z}} +
       \left ( \left . \frac{\partial e}{\partial \bar{A}} \right |_{\rho, T, \bar{Z}} + Y_e
       \left . \frac{\partial e}{\partial \bar{Z}} \right |_{\rho, T, \bar{A}} \right ) \left . \frac{\partial \bar{A}}{\partial T} \right |_{\rho, Y_e}
  \end{equation}
  where we used $\bar{Z} = Y_e \bar{A}$.  We compute
  $\partial \bar{A} / \partial{T} |_{\rho, Y_e}$ by differentiating the cubic interpolant used with the NSE table.
  \item Correct the initial guess, $T_0 \leftarrow T_0 + \delta T$.
  \item Iterate until $|\delta T| / T_0$ is small (we use a tolerance of $10^{-6}$).
\end{itemize}
This yields the $T$ and $\bar{A}$ consistent with $e^\star$.

\subsection{Integration Algorithm}

Since the NSE table gives us the instantaneous values of the state, we start by defining a procedure, ${\tt nse\_derivs}$, to estimate the time derivative of the NSE state, including the evolution due to advection:
\begin{equation}
    {\tt nse\_derivs}(\rho^s, (\rho e)^s, (\rho \alpha_l)^s ) \rightarrow [\Rb(\Uc^\prime)]^s
    %\left [d(\rho e)/dt\right ]^s, \left [d(\rho \alpha_l)/dt\right ]^s
\end{equation}
where the time-level of the state, $s$, is $n \le s \le n+1$.  ${\tt nse\_derivs}$ proceeds as:
\begin{itemize}
    \item {\em Compute the initial temperature.}
    \begin{equation}
        {\tt nse\_eos}(\rho^s, e^s, (Y_e)^s) \rightarrow T^s, \bar{A}^s
    \end{equation}
    where $e^s = (\rho e)^s/ \rho^s$ and $(Y_e)^s = (\rho Y_e)^s / \rho^s$.

    \item {\em Compute the plasma neutrino losses.}
    \begin{equation}
        \epsthermal^s = \epsthermal(\rho^s, T^s, \bar{A}^s, \bar{Z}^s)
    \end{equation}
    with $\bar{Z}^s = \bar{A}^s \cdot (Y_e)^s$ and $\bar{A}^s = (\rho \bar{A})^s / \rho^s$ from the incoming $(\rho \alpha_l)^s$. 

    \item {\em Evaluate the NSE state.}  This uses the initial thermodynamic state.
    \begin{equation}
    \nsetable(\rho^s, T^s, (Y_e)^s) \rightarrow \bar{A}^s, (\tilde{X}_k)^s \beav^s, \yedotvp^s, \dbeadt_\mathrm{weak}^s, \epsreact^s
    \end{equation}

    \item {\em Construct the initial guess of
    $[\Rb(\Uc)]^s$.}  Since we start out
    in NSE, this uses only the weak rate evolution and plasma losses:
       \begin{eqnarray}
          R^\star(\rho e) &=& N_A \rho^s \dbeadt_\mathrm{weak}^s + N_A \Delta m_{np} c^2 \rho^s \yedotvp^s -\rho^s (\epsthermal^s + \epsreact^s) \\
          R^\star(\rho Y_e) &=& \rho^s \yedotvp^s \\
          R^\star(\rho \bar{A}) &=& 0 \\
          R^\star(\rho \bea) &=& \rho^s \dbeadt_\mathrm{weak}^s
      \end{eqnarray}       

    \item {\em Evolve for a small amount of time $\tau$.}
    We pick a $\tau \ll \Delta t$ to compute
    the finite-difference approximation to the derivatives.  We'll label this new state with the superscript $\tau$:
    \begin{equation}
       \Uc^{\prime,\tau} = \Uc^{\prime, n} + \tau \Adv{\Uc^\prime}^{n+1/2} + \tau [ \Rb(\Uc^{\prime})]^\star
    \end{equation}

    \item {\em Compute the updated temperature.}
    \begin{equation}
        {\tt nse\_eos}(\rho^\tau, e^\tau, Y_e^\tau) \rightarrow T^\tau, \bar{A}^\tau
    \end{equation}

    \item {\em Evaluate the new NSE state.}
    Now we use the thermodynamics state at time $\tau$:
    \begin{equation}
    \nsetable(\rho^\tau, T^\tau, (Y_e)^\tau) \rightarrow \bar{A}^\tau, (\tilde{X}_k)^\tau, \beav^\tau, \yedotvp^\tau, \dbeadt_\mathrm{weak}^\tau, \epsreact^\tau
    \end{equation}
    It's important to note that the quantities $\bar{A}^\tau$, $(\tilde{X}_k)^\tau$ and $\bea^\tau$ returned
    from the NSE table have implicitly seen the effects
    of advection.

    \item Construct a finite-difference approximation
      to the reactive sources.  This should not include the advective contribution, so we will need to remove that.
       We start by defining
      \begin{equation}
        \label{eq:ba_no_advect}
       \widetilde{\rho \left \langle \frac{B}{A} \right \rangle } = \left [ \rho \left \langle \frac{B}{A} \right \rangle \right ]^\tau - \tau \Advss{\rho \left \langle \frac{B}{A} \right \rangle}^{n+1/2,(k)}
     \end{equation}
     and 
      \begin{equation}
        \label{eq:abar_no_advect}
       \widetilde{\rho \bar{A} } = ( \rho \bar{A})^\tau - \tau \Advss{\rho \bar{A}}^{n+1/2,(k)}
     \end{equation}   
     which removes the advective part of the update.  Then we define:
     \begin{eqnarray}
       \Delta \left (\rho \left \langle \frac{B}{A} \right \rangle \right ) &=& \widetilde{\rho \left \langle \frac{B}{A} \right \rangle } - \rho^s\left [ \left \langle \frac{B}{A} \right \rangle \right ]^\star \\
       \Delta (\rho \bar{A} ) &=& \widetilde{\rho \bar{A} } - \rho^s (\bar{A})^\star
       \end{eqnarray}
    as the change just due to the NSE evolution of the state over $\tau$.
     We then compute the sources as:
            \begin{eqnarray}
          R^s(\rho e) &=& N_A \frac{\Delta (\rho \bea)}{\tau}
                +  N_A \Delta m_{np} c^2 \rho^s \yedotvp^s
                -\rho^s (\epsthermal^s + \epsreact^s) \\
          R^s(\rho Y_e) &=& \rho^s (dY_e/dt)^s \\
          R^s(\rho \bar{A}) &=& \frac{\Delta (\rho \bar{A})}{\tau} \\
          R^s(\rho \bea) &=& \frac{\Delta (\rho \bea)}{\tau}
          \end{eqnarray}
     We note that we don't need the sources for $\tilde{X}_k$.
\end{itemize}

This completes the estimation of the reactive sources.  A typical
value of $\tau$ is $\tau = \Delta t / 20$, which is large enough for
advection to cause a change in the thermodynamic state, but overall
the convergence result appears insensitive to this choice.  This is a
runtime option in the code.

We can now do the update.  We start with the state $\Uc^{\prime,n}$, and compute the derivatives:
\begin{equation}
    {\tt nse\_derivs}(\rho^n, (\rho e)^n, (\rho \alpha_l)^n ) \rightarrow [\Rb(\Uc^\prime)]^n
\end{equation}
We then predict the state at the midpoint in time:
\begin{equation}
    \Uc^{\prime,n+1/2} = \Uc^{\prime,n} + \frac{\Delta t}{2} \Advss{\Uc^\prime}^{n+1/2} + \frac{\Delta t}{2} [\Rb(\Uc^\prime)]^n
\end{equation}
Next, we compute the derivatives at the midpoint:
\begin{equation}
    {\tt nse\_derivs}(\rho^{n+1/2}, (\rho e)^{n+1/2}, (\rho \alpha_l)^{n+1/2} ) \rightarrow [\Rb(\Uc^\prime)]^{n+1/2}
\end{equation}
and finally do the final update:
\begin{equation}
    \Uc^{\prime,n+1} = \Uc^{\prime,n} + \Delta t \Advss{\Uc^\prime}^{n+1/2} + \Delta t [\Rb(\Uc^\prime)]^{n+1/2}
\end{equation}

With the update complete, we need to compute the new mass fractions and also ensure that $\bea$ is consistent with the current thermodynamic state, so the next call will not generate energy if $\Delta t \rightarrow 0$.  Since the equation of state does not depend on $\bea$, we start by computing the updated temperature:
\begin{equation}
    {\tt nse\_eos}(\rho^{n+1}, e^{n+1}, Y_e^{n+1}) \rightarrow T^{n+1}, \bar{A}^{n+1}
\end{equation}
We then call the NSE table:
    \begin{equation}
    \nsetable(\rho^{n+1}, T^{n+1}, (Y_e)^{n+1}) \rightarrow \bar{A}^{n+1}, (\tilde{X}_k)^{n+1}, \beav^{n+1}, \yedotvp^{n+1}, \dbeadt_\mathrm{weak}^{n+1}, \epsreact^{n+1}
    \end{equation}
where, as indicated, we update $\bea$, $\bar{A}$ and $\tilde{X}_k$. 

The last piece is to update the total energy, which is not part of $\Uc^\prime$.
This requires the energy source term, which we get from the internal energy:
\begin{equation}
    (\rho \epsilon)^{n+1/2} = \frac{(\rho e)^{n+1} - (\rho e)^n}{\Delta t} - \Advss{\rho e}^{n+1/2}
\end{equation}
and then
\begin{equation}
(\rho E)^{n+1,(k)} = (\rho E)^n + \Delta t \Advss{\rho E}^{n+1/2} + \Delta t (\rho \epsilon)^{n+1/2}
\end{equation}
This completes the update.

\subsection{NSE Bailout}

Because $\rho$ and $T$ evolve during the reactions, it is possible for
a zone to start out not in NSE but evolve into NSE during the reaction
update.  When this happens, the ODE integrator may fail, because an
excessive number of timesteps is required.  To account for this during
the integration, we check to see if zone evolved into NSE, and if so,
we finish the zone update using the NSE prescription. To ensure that
this is not triggered at the very start of integration, we require a
minimum number of steps (10) to have elapsed before checking for NSE
(this is enough to get past the timestep estimation stage in VODE that
occupies the first timesteps).  After we leave the integrator, the
state is returned to the burner driver where we finish the integration
via NSE if the conditions satisfy the NSE criteria.  We optionally
apply a relaxation factor (emperically suggested to be $0.9$)
to the NSE conditions to allow a state that is close to NSE after an
integrator failure to enter NSE.  If a burn doesn't enter NSE but
still fails, then we trigger the \castro\ retry mechanism and throw
away the entire timestep and start over with a smaller $\Delta t$.

\section{Numerical Tests}

\subsection{NSE convergence test problem}

To test the coupling of the NSE table to the hydrodynamics, we run a
similar convergence test to that used in \citet{castro_sdc} and Z22 (and originally inspired by the tests in \citealt{mccorquodalecolella}), except with the thermodynamic conditions
appropriate for the matter to be in NSE.  Additionally, we add a uniform velocity field to ensure
that the advective terms have an influence on the update---this exercises the removal of the advective term done in Eqs.~\ref{eq:ba_no_advect} and \ref{eq:abar_no_advect}.  The initial conditions are:
\begin{eqnarray}
\rho &=& \rho_0 \\
T &=& \begin{cases} T_0 \left [ 1 + (\delta T) e^{-(r/\lambda)^2} \cos^6(\pi r/L) \right ] & r < r_0\\
                   T_0 & r \ge r_0 \end{cases} \\
Y_e &=& \begin{cases} (Y_e)_0 \left [ 1 + (\delta Y_e) e^{-(r/\lambda)^2} \cos^6(\pi r/L) \right ] & r < r_0 \\
                     (Y_e)_0 & r \ge r_0 \end{cases} \\
u &=& u_0 \\
v &=& v_0
\end{eqnarray}
where $r$ is the distance from the center of the domain, $L$ is the
size of the domain, $r_0 = L/2$, $u_0$ and $v_0$ are the initial $x$-
and $y$-velocities.  The value for each parameter is given in
Table~\ref{table:nse}.  We chose initial velocity values that keep the
motion subsonic, minimizing pressure fluctuations, so the hydrodynamic
slope limiters will not reduce the order of the parabolic
reconstruction and impact the convergence rate.  The peak Mach number
initially is $\mathcal{M} = 0.17$.  Our initial conditions have $Y_e$
varying between $[0.475, 0.5]$ initially. The mass fractions and
remaining composition variables are then initialized from the NSE
table.

The domain has a size $[0, L]^2$, and the timestep is fixed as:
\begin{equation}
\Delta t = 4\times 10^{-4} \left ( \frac{32}{N} \right )~\mathrm{s}
\end{equation}
where $N$ is the number of zones in each direction.  We run for 0.032
s with $N = 32$, $64$, $128$, $256$, and $512$ zones.  Since the
timestep is reduced in direct proportion to the grid spacing, this
convergence test simultaneously tests convergence in space and time.

For a pair of runs with $N$ and $2N$ zones, we coarsen the finer
resolution simulation via conservative averaging down to $N$ zones (we
call this $\phi^{(2N,\mathrm{coarsened})}$) and compute the $L_1$ norm
of the difference with the simulation run with $N$ zones, $\phi^{(N)}$
as:
\begin{equation}
    \epsilon_{N\rightarrow 2N} = \| \phi^{(N)} - \phi^{(2N,\mathrm{coarsened})} \|_1
        = \Delta x \Delta y \sum_{i,j} | \phi^{(N)}_{i,j} - \phi^{(2N,\mathrm{coarsened})}_{i,j} |
\end{equation}
We can then compute the rate of convergence, $r$, as:
\begin{equation}
    r = \log_2 \frac{\epsilon_{N\rightarrow 2N}}{\epsilon_{2N\rightarrow 4N}}
\end{equation}
Table~\ref{table:nse_sdc_methodII} shows the convergence.  For the 5 different resolutions,
we can measure 3 convergence rates.  We do this for several state variables.

We see that we get close to second order convergence for the lower
resolution runs.  At higher resolution, the size of the NSE table
seems to be the main factor affecting the convergence---with a more
widely spaced tabulation than used here, we observed worse
convergence.  This makes sense, since, as we increase the resolution,
the truncation error from the hydrodynamics will eventually get
smaller than the error from interpolating in the fixed-size table.
This test showed us the need for high-order interpolation and a dense
NSE table.  The conditions for this convergence test are
representative of those that we would find in a massive star, and we
note that the finest resolution used in the convergence test is
$0.39~\mathrm{km}$, which is much finer than the typical resolution we
would use with a star, as we show in section \ref{sec:massive_star}.

\begin{deluxetable}{lcc}
\tablecaption{\label{table:nse} NSE convergence test problem.}
\tablehead{\colhead{parameter} & \colhead{value}}
\startdata
$\rho_0$ & $5\times 10^9~\gcc$ \\
$T_0$    & $5\times 10^9~\mathrm{K}$ \\
$(\delta T)$ & $0.2$ \\
$(Y_e)_0$   & $0.5$ \\
$(\delta Y_e)$ & $-0.05$ \\
$\lambda$   & $2\times 10^7~\mathrm{cm}$ \\
$L$         & $2\times 10^7~\mathrm{cm}$ \\
$u_0$         & $10^8~\cms$ \\
$v_0$         & $10^8~\cms$ \\
\enddata
\end{deluxetable}

% run on 04-06-21
% Castro       git describe: 21.04-dirty
% AMReX        git describe: 21.03-58-g87c81d0ca
% Microphysics git describe: 21.04-1-g33a4d0fe

\begin{deluxetable}{lllllllll}
\tablecaption{\label{table:nse_sdc_methodII} Convergence ($L_1$ norm) for the NSE convergence
problem using the simplified-SDC algorithm.}
\tablehead{\colhead{field} & \colhead{$\epsilon_{32 \rightarrow 64}$} & 
           \colhead{rate} & \colhead{$\epsilon_{64 \rightarrow 128}$} & 
           \colhead{rate} & \colhead{$\epsilon_{128\rightarrow 256}$} & 
           \colhead{rate} & \colhead{$\epsilon_{256\rightarrow 512}$}}
\startdata
  $\rho$                      & $2.924 \times 10^{19}$  & 1.990  & $7.359 \times 10^{18}$  & 1.848  & $2.044 \times 10^{18}$   & 1.732  & $6.153 \times 10^{17}$  \\
 $\rho u$                    & $1.383 \times 10^{28}$  & 1.896  & $3.714 \times 10^{27}$  & 1.835  & $1.041 \times 10^{27}$   & 1.741  & $3.115 \times 10^{26}$  \\
 $\rho v$                    & $1.383 \times 10^{28}$  & 1.896  & $3.714 \times 10^{27}$  & 1.835  & $1.041 \times 10^{27}$   & 1.741  & $3.115 \times 10^{26}$  \\
 $\rho E$                    & $3.585 \times 10^{37}$  & 1.870  & $9.805 \times 10^{36}$  & 1.894  & $2.639 \times 10^{36}$   & 1.715  & $8.038 \times 10^{35}$  \\
 $\rho e$                    & $3.529 \times 10^{37}$  & 1.872  & $9.638 \times 10^{36}$  & 1.892  & $2.596 \times 10^{36}$   & 1.713  & $7.923 \times 10^{35}$  \\
 $T$                         & $7.190 \times 10^{20}$  & 2.075  & $1.706 \times 10^{20}$  & 1.921  & $4.506 \times 10^{19}$   & 1.908  & $1.200 \times 10^{19}$  \\
 $\rho X(\isotm{He}{4})$     & $2.991 \times 10^{18}$  & 1.999  & $7.484 \times 10^{17}$  & 1.914  & $1.986 \times 10^{17}$   & 1.941  & $5.172 \times 10^{16}$  \\
 $\rho X(\isotm{Cr}{48})$    & $1.645 \times 10^{19}$  & 2.188  & $3.609 \times 10^{18}$  & 2.071  & $8.589 \times 10^{17}$   & 1.987  & $2.166 \times 10^{17}$  \\
 $\rho X(\isotm{Fe}{52})$    & $4.621 \times 10^{19}$  & 1.892  & $1.245 \times 10^{19}$  & 1.849  & $3.457 \times 10^{18}$   & 1.972  & $8.814 \times 10^{17}$  \\
 $\rho X(\isotm{Fe}{54})$    & $4.251 \times 10^{20}$  & 2.230  & $9.059 \times 10^{19}$  & 2.161  & $2.026 \times 10^{19}$   & 2.158  & $4.541 \times 10^{18}$  \\
 $\rho X(\isotm{Ni}{56})$    & $4.436 \times 10^{20}$  & 2.244  & $9.365 \times 10^{19}$  & 2.160  & $2.095 \times 10^{19}$   & 2.129  & $4.789 \times 10^{18}$  \\
 $\rho X(\mathrm{n})$        & $3.835 \times 10^{14}$  & 1.710  & $1.172 \times 10^{14}$  & 1.792  & $3.383 \times 10^{13}$   & 2.035  & $8.254 \times 10^{12}$  \\
 $\rho X(\mathrm{p})$        & $8.642 \times 10^{17}$  & 2.232  & $1.840 \times 10^{17}$  & 1.845  & $5.121 \times 10^{16}$   & 1.766  & $1.506 \times 10^{16}$  \\
 $\rho Y_e$                  & $1.686 \times 10^{19}$  & 1.953  & $4.354 \times 10^{18}$  & 1.873  & $1.189 \times 10^{18}$   & 1.786  & $3.448 \times 10^{17}$  \\
 $\rho \bar{A}$              & $4.188 \times 10^{21}$  & 2.216  & $9.013 \times 10^{20}$  & 1.846  & $2.507 \times 10^{20}$   & 1.646  & $8.010 \times 10^{19}$  \\
 $\rho (B/A)$                & $2.977 \times 10^{20}$  & 1.990  & $7.493 \times 10^{19}$  & 1.881  & $2.034 \times 10^{19}$   & 1.794  & $5.864 \times 10^{18}$  \\
\enddata
\end{deluxetable}

\subsection{Massive Star}
\label{sec:massive_star}

Next we consider the evolution of a massive star, capturing
the O and Si burning shells and the evolution of the iron core on the grid.

\subsubsection{Initial Model}

\begin{figure}
\centering
\plotone{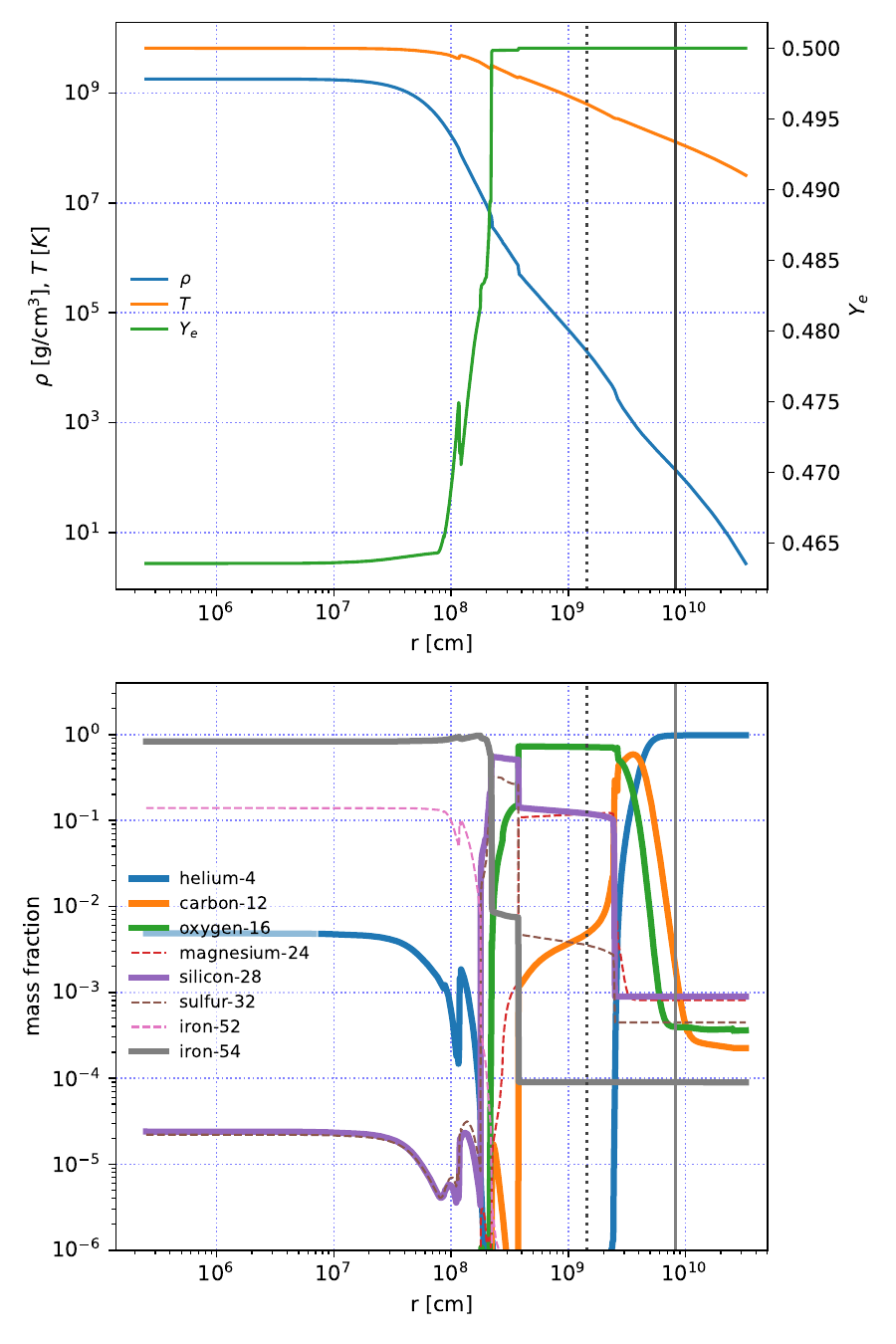}
\caption{\label{fig:initial_model} $15 M_\odot$ initial model
  (from \cite{fields:2020}) showing the structure and
  composition of the region near the core.  The dotted vertical line
  shows the radius inside of which we refine to higher resolution.
  The solid vertical line shows the radius corresponding to our domain
  width.}
\end{figure}

We use the same $15~M_\odot$ initial model as \cite{fields:2020},
generated by MESA
\citep{mesa:2011,mesa:2013,mesa:2015,mesa:2018,mesa:2019},
representing the state of the star approximately 400 s before collapse
(as determined by MESA).  At this point in time, the star's mass is about $13.3~M_\odot$.  The MESA model is interpolated onto a
uniform grid and then we reset the composition to the NSE composition
if it meets the criteria Eqs.~\ref{eq:nse_start}--\ref{eq:nse_end}.
We next need to make it consistent with our equation of state and
discrete form of hydrostatic equilibrium.  This is done by starting at
the center and using the model's $T$ and composition (including $Y_e$)
to find the $\rho$ and $p$ that satisfy HSE (in 1D spherical
coordinates) and the EOS. For zone $i$, this has the form:
\begin{equation}
    p_{i} = p(\rho) = p_{i-1}  + \frac{\Delta r}{2} (\rho + \rho_{i-1}) g_{i-1/2}
\end{equation}
where $g_{i-1/2}$ is the gravitational acceleration at the interface
between zones $i-1$ and $i$ in our initial model, computed by adding up all of the mass
interior to the interface.
We want the pressure $p_i$, and use Newton-Raphson root finding together with the equation of state to get $\rho_i$.  In the case where we are in NSE, a procedure similar to Eq.~\ref{eq:ense} and following needs to be done, but now in terms of pressure instead of energy. 
We use a 1D spacing of $\Delta r = 5~\mathrm{km}$.  Figure~\ref{fig:initial_model} shows the 
structure and composition of the initial model.  The code to remap the initial model
is freely available \citep{initial_models}.

\subsubsection{Convective Initialization}

We map the initial model onto the Cartesian grid by taking pressure,
density, composition, and $Y_e$ as inputs and interpolating them to
cell centers.  We then check if a cell is in NSE, and if so, we update
the composition and $\bea$ from the NSE table; otherwise, we compute
$\bea$ using the composition in the cell.  Finally, we call the EOS to get the temperature.  Just like with the initial model, if we are in NSE, then we need to do the EOS inversion self-consistent with the NSE table, analogous to Eq.~\ref{eq:ense}, but using the interpolated $p$ instead of $e$.  For our domain size, the
mass on the grid is $2.7~M_\odot$.

To break the symmetry,
a small velocity perturbation is added to the Si layer.  We define an amplitude:
\begin{equation}
\gamma_{l,m,n} = c^l a^m b^n (-1)^n    
\end{equation}
and phases as:
\begin{eqnarray}
    \phi^{(x)}_{l,m,n} &=& 2\pi c^l b^m a^n \\
    \phi^{(y)}_{l,m,n} &=& 2\pi b^l c^m a^n
\end{eqnarray}
and a normalization:
\begin{equation}
    \eta_{l,m,n} = (l^2 + m^2 + n^2)^{1/2}
\end{equation}
Then we define the velocity field in $x$ ($u$) and $y$ ($v$) as:
\begin{eqnarray}
   u_{i,j} &=& -A_\mathrm{pert} X(\isotm{Si}{28})\, \sum_{l=1}^3 \sum_{m=1}^3 \sum_{n=1}^3 \gamma_{l,m,n} \frac{m}{\eta_{l,m,n}}
      \cos\left(\frac{2\pi l x_i}{\lambda} + \phi^{(x)}_{l,m,n} \right )
      \sin\left(\frac{2\pi m y_j}{\lambda} + \phi^{(y)}_{l,m,n} \right ) \\
   v_{i,j} &=& A_\mathrm{pert} X(\isotm{Si}{28}) \,\sum_{l=1}^3 \sum_{m=1}^3 \sum_{n=1}^3 \gamma_{l,m,n} \frac{l}{\eta_{l,m,n}}
      \sin\left(\frac{2\pi l x_i}{\lambda} + \phi^{(x)}_{l,m,n} \right )
      \cos\left(\frac{2\pi m y_j}{\lambda} + \phi^{(y)}_{l,m,n} \right )       
\end{eqnarray}
where $\lambda$ is the characteristic scale of the fluctuations in the velocity field.
The factor of $X(\isotm{Si}{28})$ ensures that this perturbation is confined to the silicon-burning layer.
We take $a=0.5$, $b=0.7$, and $c = 0.3$, and $A_\mathrm{pert} = 5\times 10^6~\cms$, $\lambda = 10^7~\mathrm{cm}$.
The peak Mach number of this initial velocity field is $0.0016$.

When mapping from a one-dimensional model that used mixing length theory to a multi-dimensional model, there is not enough information to initialize the velocity field.  This means that as soon as the simulation begins, the burning at the base of the convective layer will release lots of energy that is not carried away, resulting in a nonlinear feedback that can cause problems (for example, thermonuclear runaways).  To help mitigate this problem, we use a special initialization procedure to help establish a convective field.  This is the {\em drive initial convection} feature in \castro.  The
basic algorithm is:
\begin{itemize}
    \item Allow star to evolve for a time $\tau_\mathrm{reset}$.  During this time, calls to the nuclear reaction network will use the temperature interpolated from the initial model instead of the temperature on the grid.  This breaks the nonlinear runaway that can occur from the burning happening "in-place" since there is no convective field to carry the energy away.  Note: the SDC-NSE update,
    described in section \ref{sec:sdc-nse}, differs slightly
    during this phase, since the temperature is held fixed.
    
    \item After the $\tau_\mathrm{reset}$ has elapsed, we reinitialize the initial thermodynamic state (by again mapping directly from the initial model), but leave the velocity field unchanged from what is on the grid.
    \item We repeat this evolve-remap process every $\tau_\mathrm{reset}$.
    This will slowly build up a convective
    velocity field  that respects the initial model.
    \item After a time $\tau_\mathrm{drive}$,
    we stop this initialization process and evolve as normal.
\end{itemize}    
A different solution to this problem was used in the anelastic
calculations of core convection in Chandrasekhar mass white dwarfs by \citet{kuhlen:2006}, where
the energy generation was expressed as a power law
in terms of the background temperature, with linear perturbations for the duration of the simulation.  That works well for the ``simmering'' they modeled. 
 In our case, we need to capture the nonlinearity of the reactions, and our method allows for that once the convective field is established.
For these simulations, unless otherwise noted, we use $\tau_\mathrm{reset} = 2~\mathrm{s}$ and $\tau_\mathrm{drive} = 50~\mathrm{s}$. 

%This large value of $\tau_\mathrm{drive}$ was chosen to give the oxygen-shell convection a chance to establish.

\subsubsection{Simulations}

\begin{deluxetable*}{lrrllllll}
\tablecaption{\label{table:simulations} Parameters for our simulation suite.}
\tablehead{\colhead{name} &
           \colhead{domain size} &
           \colhead{coarse grid} &
           \colhead{refinement} &
           \colhead{CFL} &
           \colhead{$\tau_\mathrm{drive}$} &
           \colhead{$\rho_\mathrm{sponge}$} &
           \colhead{note} \\
           \colhead{} & \colhead{(km)} & \colhead{} & \colhead{(num : jumps)} & \colhead{} & \colhead{(s)} & \colhead{$\gcc$}
}
\startdata
{\tt ms} & $8192 \times 16384$
     & $1024 \times 2048$ 
     & 2: $2\times, 2\times$
     & 0.4
     & 50
     & $10^3$
     & \\
{\tt ms-100} & $8192 \times 16384$
     & $1024 \times 2048$ 
     & 2: $2\times, 2\times$
     & 0.4
     & 100
     & $10^3$
     & \\
{\tt ms-cfl2} & $8192 \times 16384$
     & $1024 \times 2048$ 
     & 2: $2\times, 2\times$
     & 0.2
     & 50
     & $10^3$
     & \\
{\tt ms-cfl8} & $8192 \times 16384$
     & $1024 \times 2048$ 
     & 2: $2\times, 2\times$
     & 0.8
     & 50
     & $10^3$
     & \\
{\tt ms-sp} & $12288 \times 24576$
     & $768 \times 1536$ 
     & 3: $2\times, 2\times, 2\times$
     & 0.4
     & 50
     & $10^2$
     & \\
{\tt ms-enu} & $8192 \times 16384$
     & $1024 \times 2048$ 
     & 2: $2\times, 2\times$
     & 0.4
     & 50
     & $10^3$
     & includes $\epsreact$\\
{\tt ms-noburn} & $8192 \times 16384$
     & $1024 \times 2048$ 
     & 2: $2\times, 2\times$
     & 0.4
     & 50
     & $10^3$
     & no pert; no reactions\\ 
\enddata
\end{deluxetable*}

We run 7 simulations, each with slightly different parameters to assess the sensitivity of the evolution to the problem setup.  Table ~\ref{table:simulations} summarizes
the simulations.  The default simulation, {\tt ms}, uses a domain that is 
$8.192\times 10^9~\mathrm{cm} \times 1.6384\times 10^9~\mathrm{cm}$ in a 2D axisymmetric ($r$-$z$) geometry.
The base grid is $1024\times 2048$ zones and 2 levels of refinement are used, each jumping by a factor of 2, giving a fine-grid resolution of 20 km.  This is slightly coarser than the 16 km resolution of \cite{couch:2015}.  Refinement is done by tagging all cells that have a density larger than $10^4~\gcc$, encompassing most of the O shell.
Figure~\ref{fig:initial_model} shows the extent of the refined region and the domain.  Subcycling is used to advance coarse zones at
larger timesteps than fine zones.
The gravitational acceleration is computed as a simple monopole, with the source term treated conservatively (as described in \citealt{wdmergerI}).  Integration is done with
CFL = 0.4, without any reaction-based timestep limiter.   Evolving to $t - \tau_\mathrm{drive} = 300 s$ takes 98755 coarse timesteps for {\tt ms}.

The $r = 0$ boundary is reflecting, enforcing the axisymmetric geometry.  For
the three other boundaries, we use a simple zero-gradient outflow.  However,
this alone is not sufficient to keep the star stable in hydrostatic equilibrium when when there are no reactions.  In the work of \citet{chatzopoulos:2016}, a diode boundary condition was used---allowing material to exit the domain but not enter.  Instead, we use a sponging term to the momentum equation, following the same prescription of \cite{eiden:2020}.  This turns on at a density of $\rho_\mathrm{sponge}$ and comes into full effect when the density drops by an order of magntiude.  For the {\tt ms} simulation, we use $\rho_\mathrm{sponge} = 10^3~\gcc$.  This is outside of the O shell.  We'll explore the effect of this choice in the {\tt ms-sp} simulation.

The contribution to the energy generation from weak-rate neutrinos, $\epsreact$, is disabled---as we'll see shortly, including this term leads to an almost instantaneous collapse, suggesting a mismatch between the weak rate neutrino physics in the MESA model and what we've modeled here.  The initial model we are using was run in MESA with the {\tt aprox21} network, which does not include all of these neutrinos.  Consistency between
the nuclear physics used in the 1D model and that used here will be explored in a follow-on paper.  The convection is driven for $\tau_\mathrm{drive} = 50~\mathrm{s}$, after which the problem is left to evolve.

All simulations are run on the NERSC Perlmutter machine using 8 nodes, and running completely on the 4 NVIDIA A100 GPUs / node using CUDA
through the C++ performance portability framework enabled in \amrex~\citep{amrex_joss}.
Evolving in 2D for 200 seconds takes about 244 node-hours.

\begin{figure}
    \centering
    \plotone{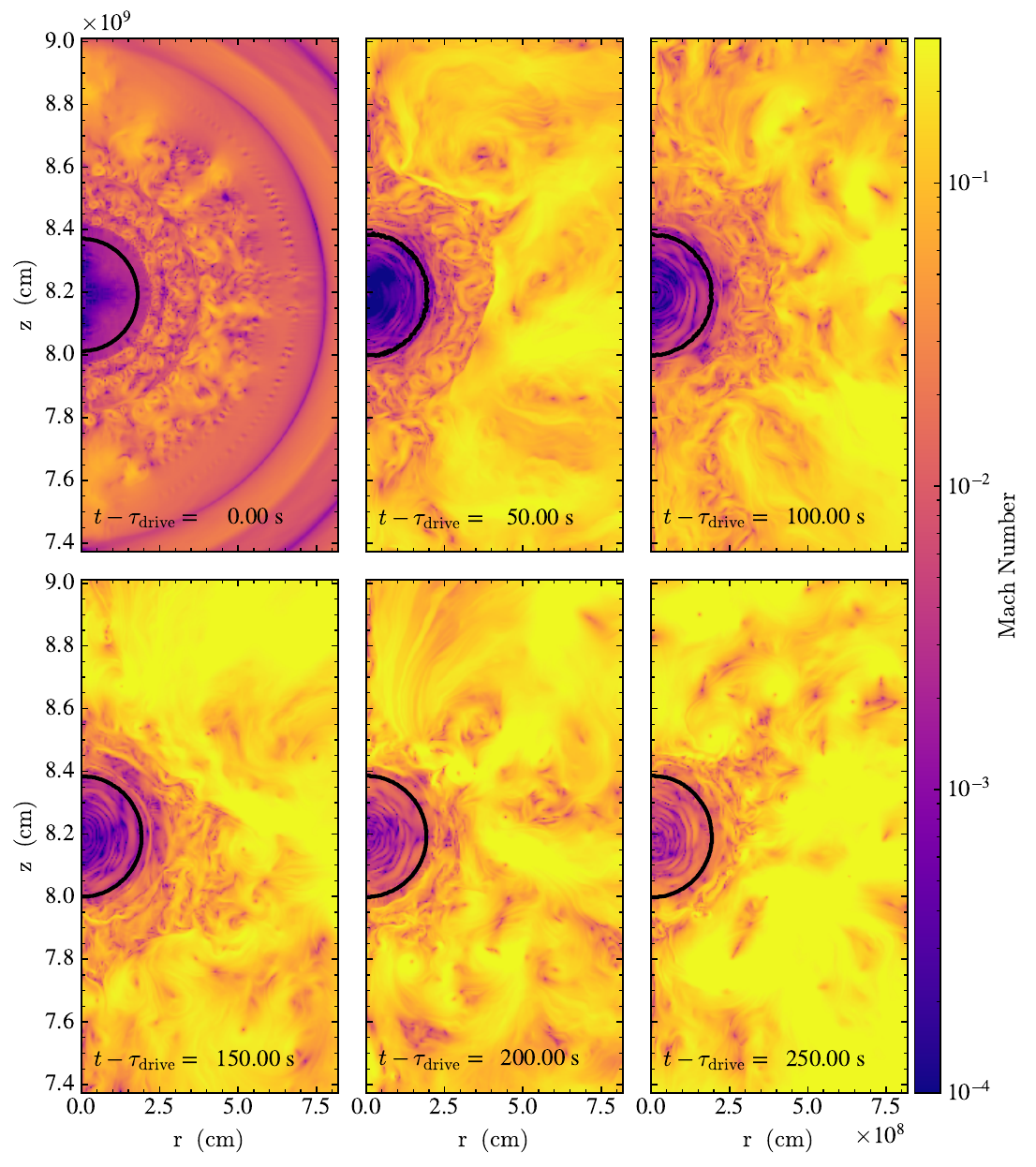}
    \caption{\label{fig:mach_sequence} Mach number sequence for the
      {\tt ms} simulation focusing on the region near the core.  The
      first frame is just after the drive initial convection phase. The black contour separates the region
      where the NSE table is used from the region where the full
      network is integrated.}
\end{figure}

\begin{figure}
    \centering \plotone{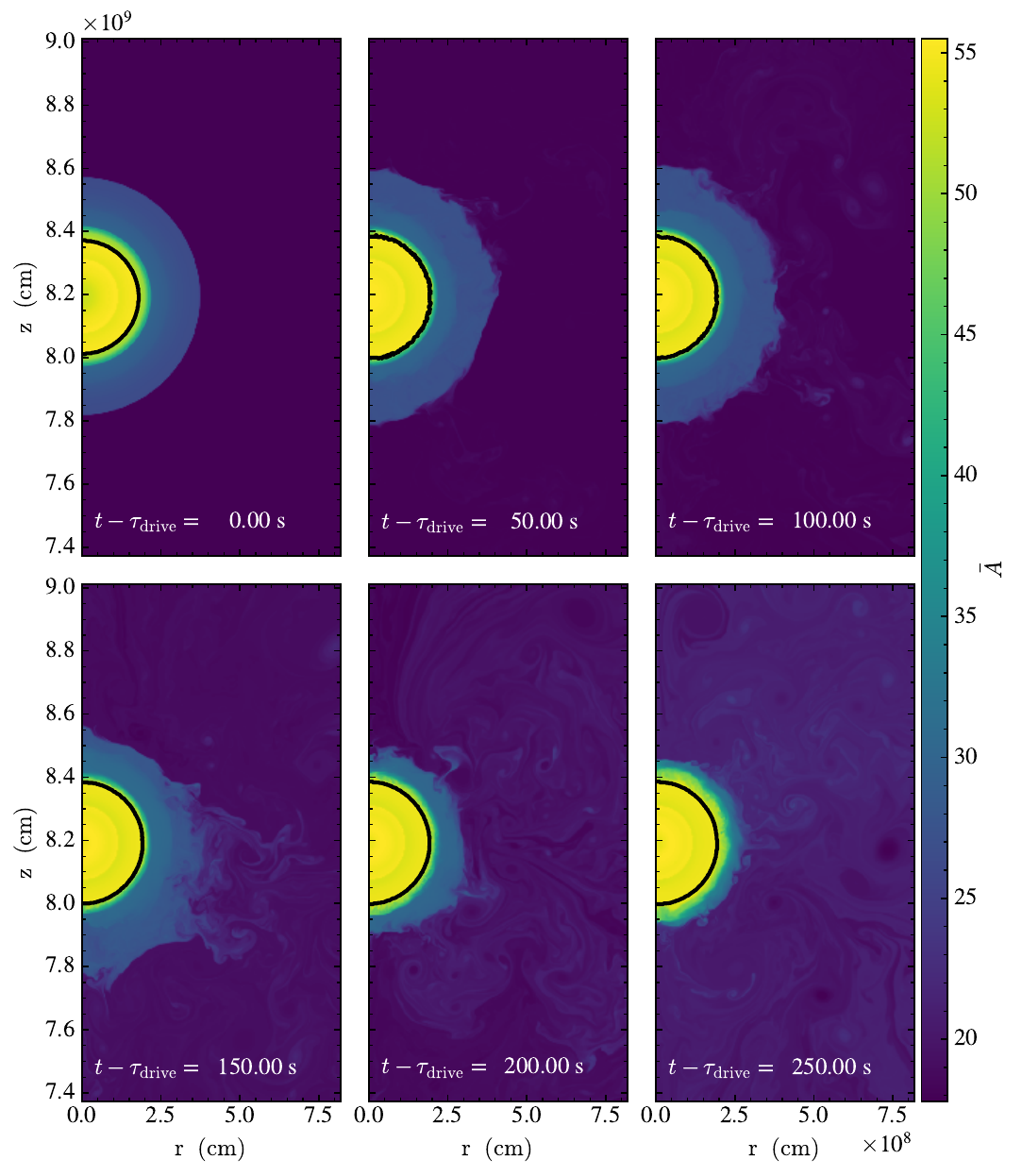}
    \caption{\label{fig:abar_sequence} Time sequence of $\bar{A}$ for
      the {\tt ms} simulation, focusing on the region near the
      core. The black contour separates the region where the
        NSE table is used from the region where the full network is
        integrated.}
\end{figure}

\begin{figure}
    \centering \plotone{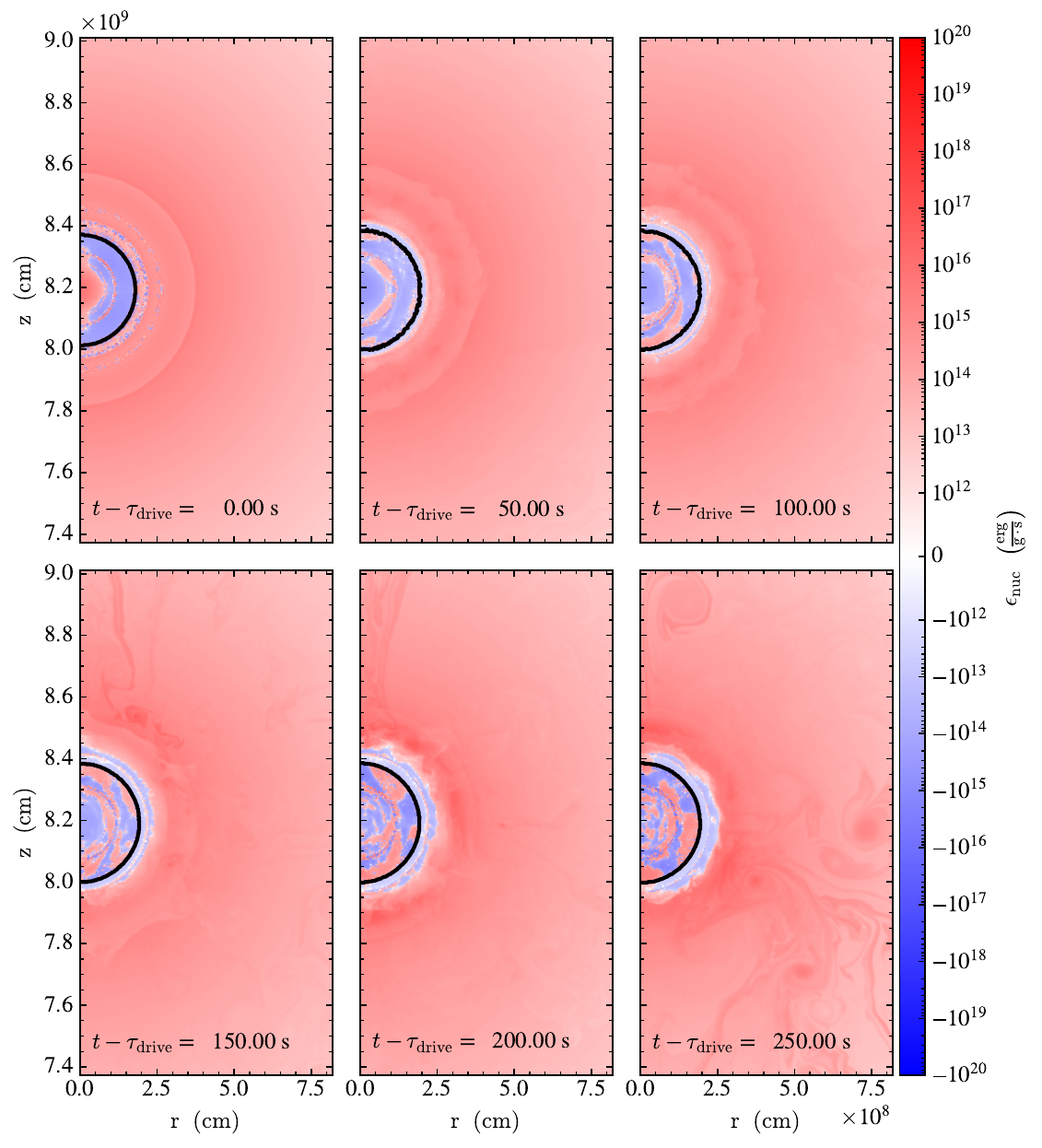}
    \caption{\label{fig:enuc_sequence} Time sequence of the energy
      generation rate for the {\tt ms} simulation focusing on the
      region near the core.  The black contour separates the region
      where the NSE table is used from the region where the full
      network is integrated.}
\end{figure}

Figures~\ref{fig:mach_sequence}--\ref{fig:enuc_sequence} show the evolution of the
{\tt ms} simulation in the region close to the core (only 10\% of the domain in each dimension is shown).  The first panel is immediately after the convective initialization procedure ends.  As time evolves, we see well-defined convection in the Si shell, but at the later times, as the convection in the O shell becomes more vigorous, the shell begins to erode.  This is likely an effect of the two-dimensional nature of the simulation.  Since our focus for this paper is the time-integration algorithm, we do not explore 3D here.
 
\begin{figure}
    \centering
    \plotone{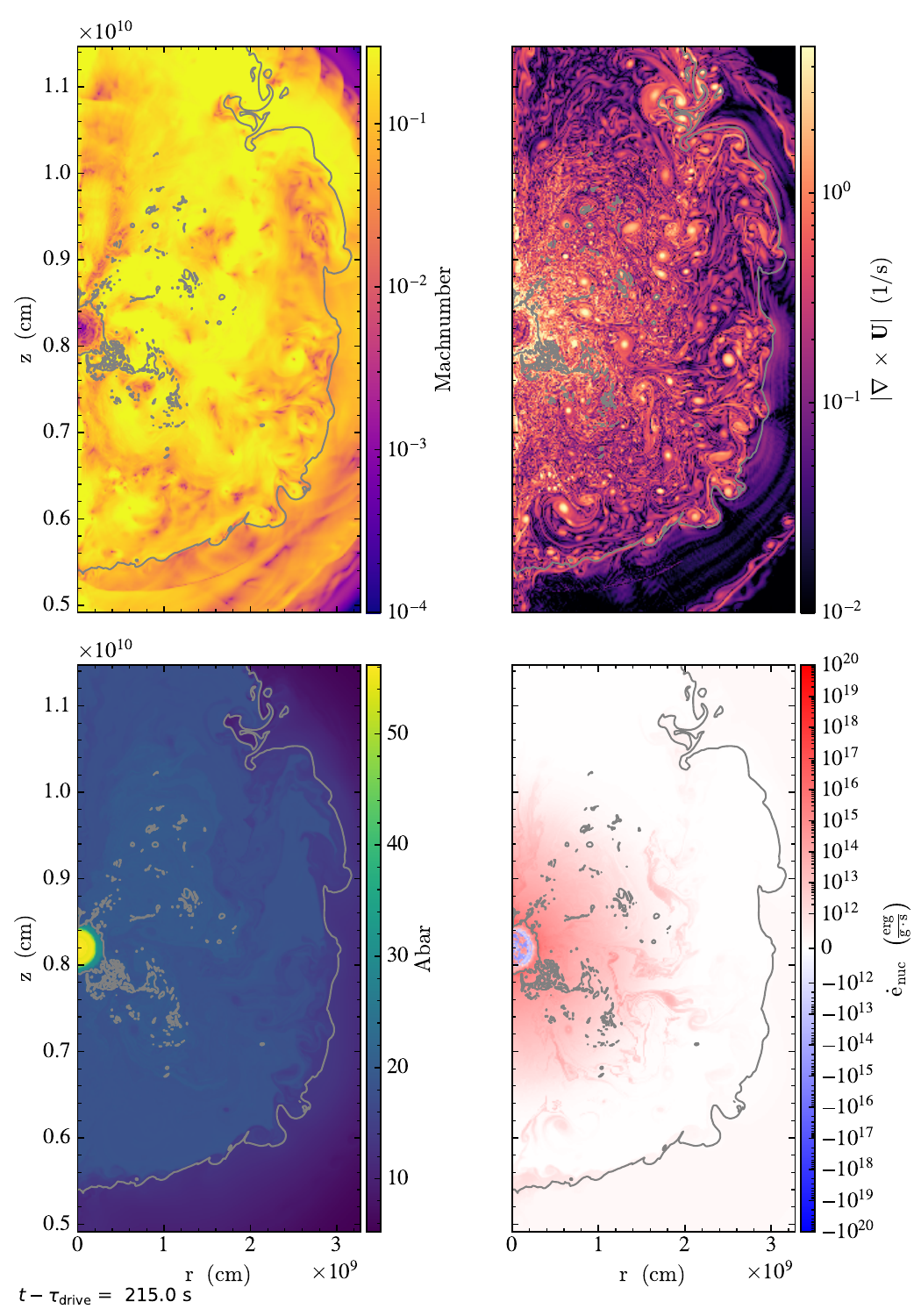}
    \caption{\label{fig:zoom_out} The state of the convection in the
      {\tt ms} simulation in a large region of the star, 200~s after
      the drive initial convection.  The full O shell is seen on this
      scale.  The gray contour marks an $\isotm{O}{16}$ mass fraction
      of $0.5$.}
\end{figure}

Zooming out, Figure~\ref{fig:zoom_out} shows the state on a much
larger scale (four times larger than
  Figures~\ref{fig:mach_sequence}--\ref{fig:enuc_sequence}).
  A gray contour is added that marks $X(\isotm{O}{16}) = 0.5$,
  which both demonstrates the extent of the oxygen shell and that
  there is mixing near the core resulting in pockets depleted of
  oxygen.  We also clearly see the large scale vortices in
the O shell that dominate the flow, and note that the Mach
number can get quite high there.  Again, we expect much of this
behavior to be an artifact of the 2D geometry.

\begin{figure}[t]
    \centering
    \plottwo{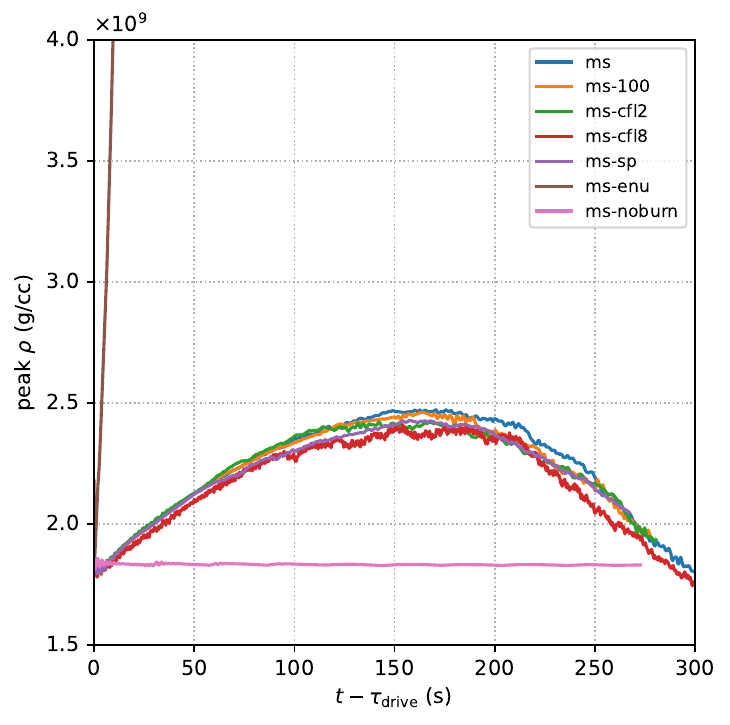}{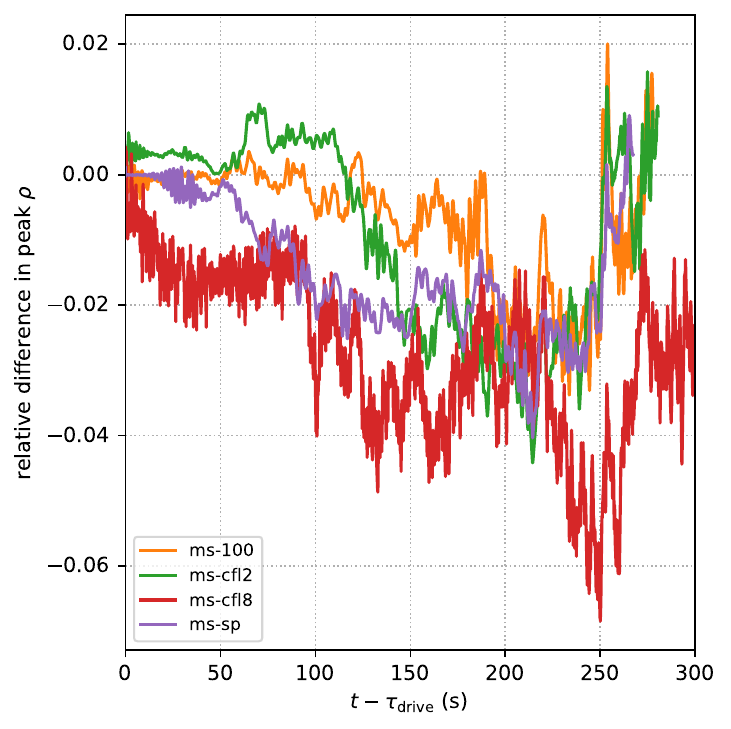}
    \caption{\label{fig:peak_rho} (left) Peak density vs.\ time since the
      drive initial convection process ended for the 7 different
      simulations.  (right) Relative difference in peak density with the
      {\tt ms} simulation for several cases.}
\end{figure}

\begin{figure}[t]
    \centering
    \plottwo{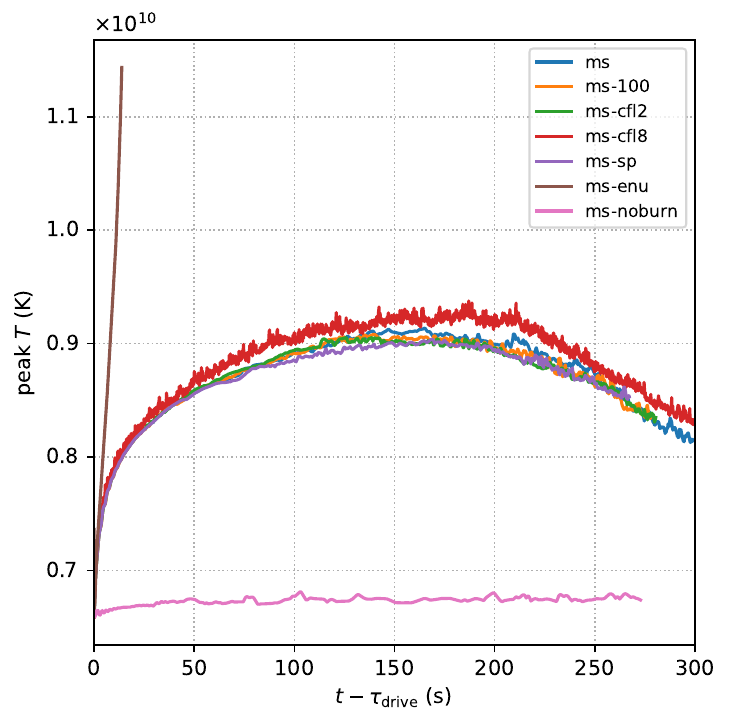}{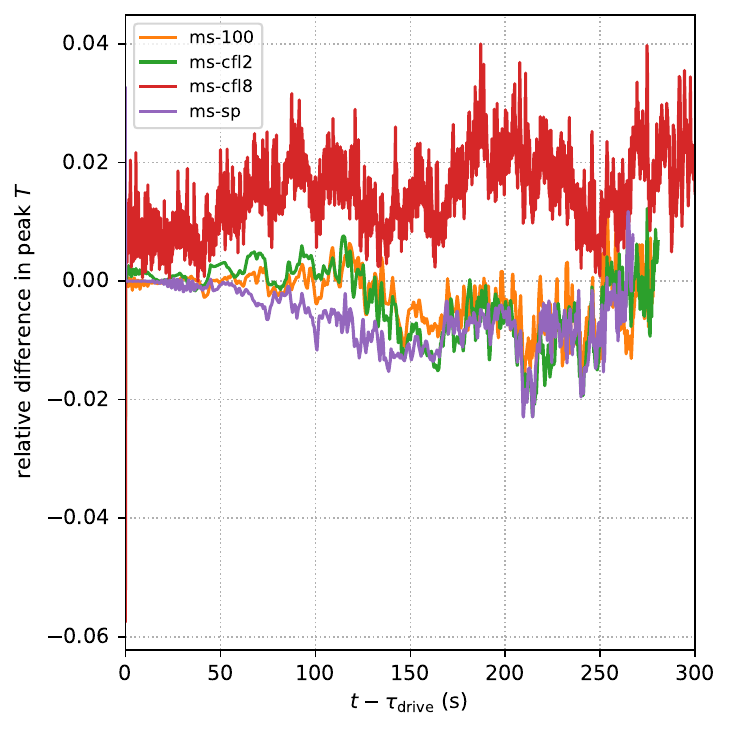}
    \caption{\label{fig:peak_T} (left) Peak temperature vs.\ time since the
      drive initial convection process ended for the 7 different
      simulations.  (right) Relative difference in peak temperature with the
      {\tt ms} simulation for several cases.}
\end{figure}

\begin{figure}[t]
    \centering
    \plottwo{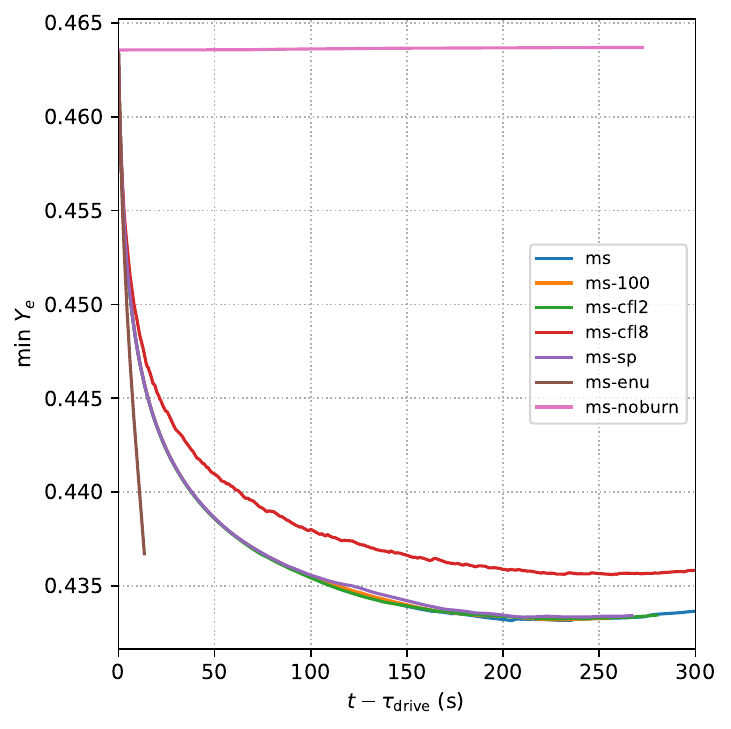}{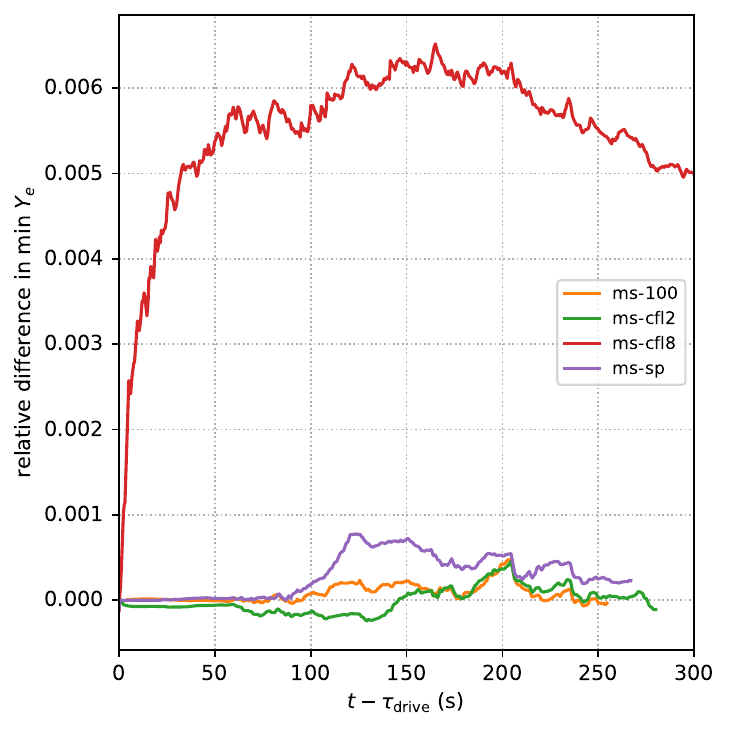}
    \caption{\label{fig:min_ye} (left) Minimum $Y_e$ vs.\ time since the
      drive initial convection process ended for the 7 different
      simulations.  (right) Relative difference in minimum $Y_e$ with the
      {\tt ms} simulation for several cases.}
\end{figure}

\begin{figure}[t]
    \centering
    \plottwo{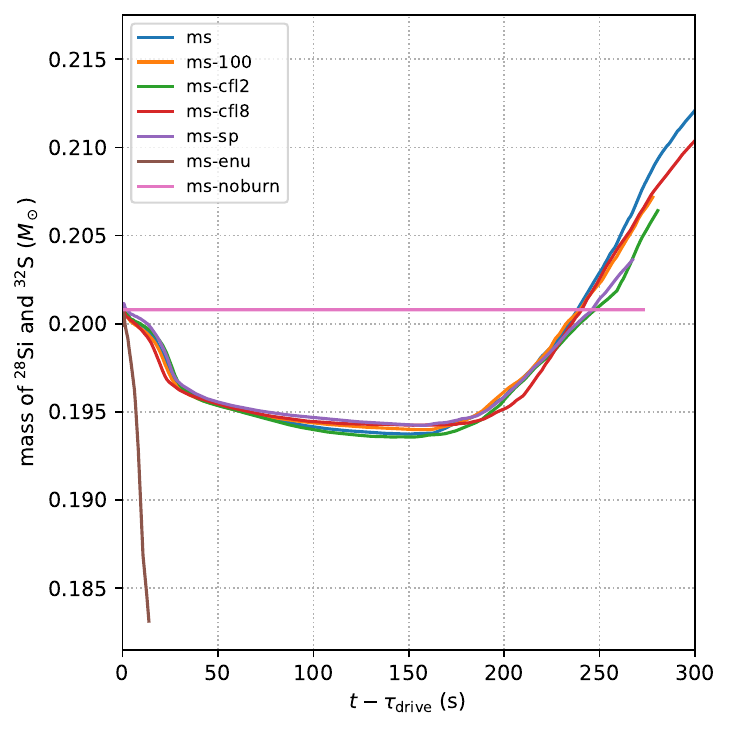}{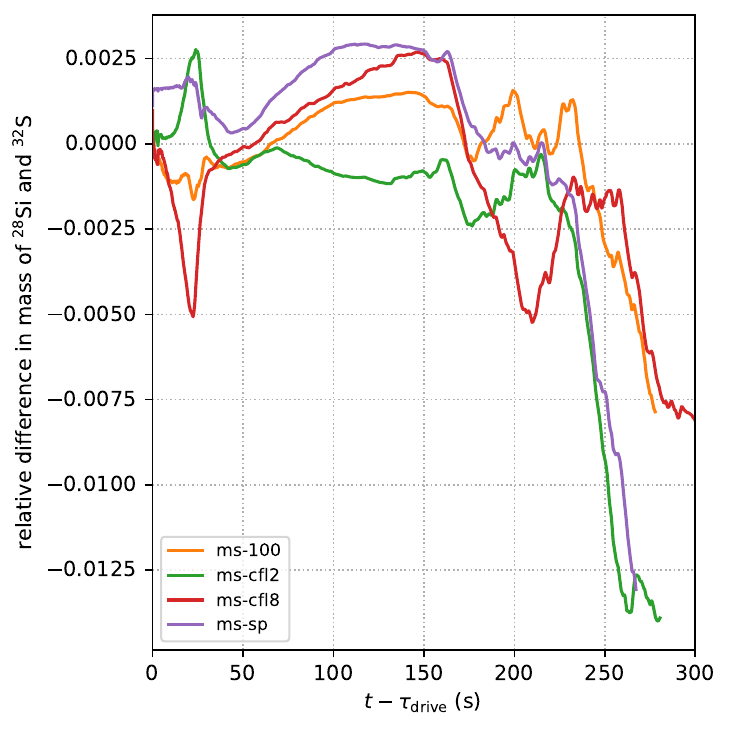}
    \caption{\label{fig:si_mass} Sum of the $\isotm{Si}{28}$ and
      $\isotm{S}{32}$ mass vs.\ time since the drive initial convection
      process ended for the 7 different simulations.  (right) Relative
      difference in the total $\isotm{Si}{28}$ and
      $\isotm{S}{32}$ mass with the {\tt ms} simulation for
      several cases.}
\end{figure}

Next we look at how the integral properties of the simulations differ
as we adjust some model parameters.  The {\tt ms-100} simulation
doubles $\tau_\mathrm{drive}$ from $50~\mathrm{s}$ to
$100~\mathrm{s}$.  The {\tt ms-cfl2} and {\tt ms-cfl8} simulations
explore CFL numbers of 0.2 and 0.8, respectively.  The {\tt ms-sp}
simulation uses a domain that is 50\% larger (in each dimension) and
drops $\rho_\mathrm{sponge}$ from $10^3~\gcc$ to $100~\gcc$---this
allows us to understand how well our outer boundary works.  This also
means that the evolution of the full C shell is now captured on the
grid.  Figure~\ref{fig:peak_rho} shows the evolution of the maximum
density on the grid as a function of time for the different
simulations as well as the difference with respect to the {\tt ms}
simulation for {\tt ms-100}, {\tt ms-cfl2}, {\tt ms-cfl8}, and {\tt
  ms-sp}.  For those four simulations, the central density slowly
increases in time, reaches a peak around 150 s, and then begins to
decrease.  This peak corresponds to the point shown in
Figure~\ref{fig:abar_sequence} where the O convection begins to erode
the Si layer.  Again we note that this is very likely a 2D effect.
The largest relative difference in density is less than 1\% for these
simulations, demonstrating strong agreement as we vary the simulation
parameters. Figure~\ref{fig:peak_T} shows the peak temperature
evolution and difference with respect to {\tt ms}, and we see
the same trend for this group of simulations,with the maximum
  difference around 0.5\%.  Figure~\ref{fig:min_ye} shows the minimum
$Y_e$ on the grid.  All simulations show a decrease in time, and none
fall below $Y_e = 0.43$, which is the boundary of our table.
The comparison with {\tt ms} shows that the largest difference
  in the global $Y_e$ minimum is less than 1\%.  Finally,
Figure~\ref{fig:si_mass} shows the total mass of $\isotm{Si}{28}$ and
$\isotm{S}{32}$.  We see the mass of these elements decreasing
initially, but again, near 150 s, the trend inverts and we begin
producing more of these nuclei.  This is consistent with the idea that
the strong 2D O-shell convection begins disrupting the Si shell,
bringing fresh O into hot regions and increases these abundances.
Once again, the comparison with the {\tt ms} simulation shows
  strong agreement across the variations, with the largest difference
  around 1\%.  The key takeaway from these figures is that our
evolution is largely insensitive to the domain size / sponging, CFL,
and initialization.

Finally the {\tt ms-enu} simulation enables the $\epsreact$ term in the energy generation in the NSE region.   We see that the density increases
dramatically right from the start, leading to a collapse.
This suggests that the MESA initial model, which used {\tt aprox21}, did not capture all of these neutrino losses.

One more simulation is done---{\tt ms-noburn} disables the initial velocity perturbation and turns off all reactions, to serve as a check that the star remains in HSE on our grid for long timescales.  As noted in \cite{mocak:2010}, the temperature gradient from the initial model can drive convective motions in multi-dimensional simulations, so disabling reactions does not mean no motion on the grid.  But the run with {\tt ms-noburn} should not show any largescale structure evolution.  We see that the density remains nearly constant in this simulation, showing that our star is stable in the absence of energy generation.  Interestingly, if we do this same simulation without the sponge, the star eventually collapses as material flows through the outer boundary.  This demonstrates that our treatment of the outer boundary with a sponge is essential.

\section{Summary}

The focus of this paper was the development of an accurate
time-integration strategy for reactive flows that enter nuclear
statistical equilibrium.  We showed that our new simplified spectral
deferred corrections scheme for coupling hydrodynamics, reactions, and
NSE is second-order accurate and provides an effective simulation
method for modeling convection in massive stars, including the
evolution of the core on the grid.  Furthermore, the time-integration
algorithm allows for evolution on the hydrodynamics timescale, and our
tests investigating the timestep size and initialization procedure
showed that the simulations are robust, with only small (< 10\%)
variations.  Furthermore, with GPU acceleration, the new algorithm is
very efficient on today's supercomputers.  This entire simulation
methodology is freely available in the AMReX-Astro github
organization\footnote{https://github.com/AMReX-Astro/}.

Additionally, a procedure for driving the initial convection was shown to help establish
a robust convective field for the initial model.  It should be possible to do
this at a coarser resolution and then add a refinement level (or more) for the last
few $\tau_\mathrm{drive}$ periods to get the benefits of this initialization without a large computational cost.

This work is the basis for a study of convection in massive stars in
3D.  Currently we are fully refining all regions where the density is
greater than $10^4~\gcc$.  This corresponds to a radius of about
20,000 km.  In \citet{couch:2015}, only the inner 2,500 km was fully
refined.  To reduce the computational cost for 3D, we will need to
relax our refinement with radius compared to the 2D models presented
here.  Initial estimates show that a 3-level simulation, evolving
$\rho > 3\times 10^5~\gcc$ at 20 km (corresponding to $r <
4700~\mathrm{km}$), $\rho > 2\times 10^3~\gcc$ at 80 km, and the base
grid at 320 km resolution (each jump is $4\times$) will need
approximately 40,000 node hours on the OLCF Frontier machine to evolve
for 350 s (including the 50 s of convective initialization).  This
estimate uses HIP to offload the compute kernels to the AMD Instinct
MI250X GPUs, similar to how we use CUDA with NVIDIA GPUs at NERSC.  We
expect to bring this cost down with further optimization on AMD GPUs.

It is straightforward to adapt this to the case where we compute the NSE abundance from the ongrid network itself instead of using a table.  The same time integration algorithm will apply when the entire network is in NSE, which can be detected using the same procedures described, for example, in \citet{Kushinr}.  This can be applied to the double detonation model in Type Ia supernovae and benefit from the second-order accuracy demonstrated here.

This same overall integration framework can apply to our low-Mach
number stellar hydrodynamics code \maestroex~\citep{maestroex}.
We will explore this in the future to possibly allow for longer-term evolution beginning at earlier times if we find that the Mach numbers are low enough.  This was used previously to model H core convection in massive stars \citep{ms_cc}. \maestroex\ and \castro\ share the same AMReX-Astrophysics \microphysics\ package.

\section*{acknowledgements}
 \castro\ and the AMReX-Astrophysics suite are freely available at
\url{http://github.com/AMReX-Astro/}.  All of the code and
problem setups used here are available in the git repo: the convergence
test is in {\tt Castro/Exec/reacting\_tests/nse\_test} and 
the massive star convection problem is in {\tt Castro/Exec/science/massive\_star}.  The script used to
generate the NSE grid is in {\tt Microphysics/nse\_tabular}.
We thank Carl Fields for sharing the initial model with us.
We thank Stan Woosley for helpful discussions about the NSE table used in
the work of \citet{ma:2013}.  The work at
Stony Brook was supported by DOE/Office of Nuclear Physics grant
DE-FG02-87ER40317.  This material is based upon work supported by the
U.S. Department of Energy, Office of Science, Office of Advanced
Scientific Computing Research and Office of Nuclear Physics,
Scientific Discovery through Advanced Computing (SciDAC) program under
Award Number DE-SC0017955. This research used resources of the National Energy Research Scientific Computing Center (NERSC), a Department of Energy Office of Science User Facility using NERSC award NP-ERCAP0027167.  This research used resources of the Oak Ridge Leadership Computing Facility at the Oak Ridge National Laboratory, which is supported by the Office of Science of the U.S. Department of Energy under Contract No. DE-AC05-00OR22725.

\software{\amrex\ \citep{amrex_joss},
          \castro\ \citep{castro, castro_joss},
          GNU Compiler Collection (\url{https://gcc.gnu.org/}),
          Linux (\url{https://www.kernel.org}),
          matplotlib (\citealt{Hunter:2007},  \url{http://matplotlib.org/}),
          NumPy \citep{numpy,numpy2,numpy2020},
          \pynucastro\ \citep{pynucastro,pynucastro2},
          python (\url{https://www.python.org/}), 
          SciPy \citep{scipy,scipy2},
          SymPy \citep{sympy},
          valgrind \citep{valgrind},
          yt \citep{yt}
         }

\facilities{NERSC, OLCF}

\appendix

\section{Constructing the NSE Table}
\label{app:nse}

We use \pynucastro\ \citep{pynucastro,pynucastro2} to construct the NSE
table, making use of the features in the latest version \citep{pynucastro21}.  
We use the weak rates from \cite{langanke:2001} to be consistent with the work from \cite{seitenzhal:2009,ma:2013}.  By using
a completely open, reproducible framework built from \pynucastro,
we can regenerate the table easily as newer rate compilations
are added to \pynucastro, for example the those of \citet{giraud:2022}.  The solution procedure follows that of \citet{seitenzhal:2009}, and we note that we assume that the nuclei behave as an ideal gas, consistent with our equation of state and a good approximation for the densities we model here.

Nuclear masses are taken from the atomic mass evaluation (AME) 2016 database \citep{ame2016,ame2016b} and spins from \citet{nubase2020}.
The temperature-dependent partition functions come from
\citet{rauscher:1997,rauscher:2003}.  Finally, we use the 
Coulomb screening implementation from \citet{chabrier_screening}.

We setup a network with 96 nuclei: $\mathrm{n}$, $\mathrm{p}$, $\mathrm{d}$, $\isotm{He}{3\mbox{--}4}$, $\isotm{C}{12}$,
$\isotm{N}{13\mbox{--}14}$, $\isotm{O}{16}$, $\isotm{F}{18}$, $\isotm{Ne}{20\mbox{--}22}$, $\isotm{Na}{23}$, $\isotm{Mg}{24}$, $\isotm{Al}{27}$, $\isotm{Si}{28}$, $\isotm{P}{31}$, $\isotm{S}{32}$, $\isotm{Cl}{35}$, $\isotm{Ar}{36}$, $\isotm{K}{39}$, $\isotm{Ca}{40,45\mbox{--}48}$, $\isotm{Sc}{43,45\mbox{--}49}$, $\isotm{Ti}{44\mbox{--}52}$, $\isotm{V}{47\mbox{--}54}$, $\isotm{Cr}{48\mbox{--}56}$, $\isotm{Mn}{51\mbox{--}58}$, $\isotm{Fe}{52\mbox{--}60}$, $\isotm{Co}{54\mbox{--}61}$, $\isotm{Ni}{56\mbox{--}65}$, $\isotm{Cu}{59}$, $\isotm{Zn}{60}$.
This  selection of nuclei is smaller than that of \citet{ma:2013} and \citet{seitenzhal:2009}, in part because we are restricting ourselves to only those nuclei with well-measured spins.  
We will explore a wider range of nuclei and adopt a more recent AME dataset in a later paper.

Using \pynucastro, we create an {\tt NSENetwork} with our set of nuclei.  This can then solve the NSE constraint equations and give
us the equilibrium state.
We then create 
a grid of density, temperature,
and electron fraction.  In all, we use 101 temperature points, 
equally spaced in logarithm between $T = 10^{9.4}~\mathrm{K}$ and $T = 10^{10.4}~\mathrm{K}$, 61 density points equally spaced in logarithm between $\rho = 10^7~\gcc$ and $\rho = 10^{10}~\gcc$, and 29 $Y_e$ points, linearly spaced between $Y_e = 0.43$ and $Y_e = 0.5$.  These correspond to spacings:
\begin{eqnarray}
    \Delta \log (T / 1~\mathrm{K}) &=& 0.01 \\
    \Delta \log (\rho / 1~\gcc) &=& 0.05 \\
    \Delta \log Y_e &=& 0.0025
\end{eqnarray}
This fine spacing was needed to get the second-order convergence in our test problem.
At each grid point,
we compute the NSE state, $X_k^\mathrm{NSE}$.
The NSE state will evolve as electron captures
and $\beta$-decays take place and alter $Y_e$.
We can compute this evolution by calling the 
RHS of the network with $X_k^\mathrm{NSE}$:
\begin{equation}
\omegadot_{k,\mathrm{weak}} = \mathcal{F}_\mathrm{weak}\{\omegadot_k (\rho, T, X_k^\mathrm{NSE})\}
\end{equation}
where $\mathcal{F}_\mathrm{weak}\{\}$ represents a filter on the rates that only considers the tabulated weak rates.
In \pynucastro, this filtering is done by passing a {\tt rate\_filter} function to {\tt RateCollection.evaluate\_ydots()} 
that only acts on the tabulated weak rates.

We store the mean molecular weight of the NSE state:
\begin{equation}
\bar{A} = \sum_k \dfrac{X_k^\mathrm{NSE}}{A_k}
\end{equation}
the average binding energy per nucleon of the NSE state:
\begin{equation}
\left \langle \frac{B}{A} \right \rangle = \sum_k \dfrac{X_k^\mathrm{NSE} B_k}{A_k}
\end{equation}
the evolution of the electron fraction:
\begin{equation}
\dot{Y_e} = \sum_k \dfrac{\omegadot_{k, \mathrm{weak}} Z_k}{A_k}
\end{equation}
the evolution of the binding energy / nucleon due to weak
reactions:
\begin{equation}
    \dbeadt_\mathrm{weak} = -\bar{A}^2 \sum_k \frac{\omegadot_{k,weak} B_k}{A_k}
\end{equation}
and the neutrino loss rate: 
\begin{equation}
\epsreact = \sum_{r \in \mathrm{rates}} Y_r {\epsilon_\nu}_r(\rho Y_e, T)
\end{equation}
where $Y_r$ is the molar abundance of the nucleus that participates in the decay or capture in rate $r$.

Finally, we then bin the full set of nuclei down to the
19 we store on the grid:
\begin{equation}
\tilde{X_k} = \mathrm{bin} (\{X_j^\mathrm{NSE}\})
\end{equation}
This binning is handled automatically in the \pynucastro\
{\tt Composition.bin\_as()} function and seeks to add each NSE nucleus's mass fraction to the corresponding {\tt aprox19} mass fraction by finding
the lightest {\tt aprox19} nucleus with an atomic mass greater than or equal to the mass of the NSE nucleus.  In the case of multiple matches, we then consider the proton number.  We make one exception here, as in \citet{ma:2013}, and do not add anything to \isot{Ni}{56} except for the NSE abundance of \isot{Ni}{56}.  This is due to the role of \isot{Ni}{56} in lightcurves (although not really needed for this application).
We also note that the {\tt aprox19} network includes both \isot{H}{1} and protons with the former participating
in H burning and the latter in photodisintegration links in the heavy elements.  This is done to reduce the coupling in the network and simplify the linear algebra.  For our NSE state, we only map into the protons.

%======================================================================
% References
%======================================================================

\bibliographystyle{aasjournal}
\bibliography{ws}

\end{document}